\documentclass[12pt]{article}
\usepackage{latexsym}
\usepackage[dvips]{graphicx}
\newcommand{\mev}{\,{\rm MeV}}
\newcommand{\gev}{\,{\rm GeV}}
\newcommand{\gevcc}{\,{\rm GeV}/c^2}
\newcommand{\mic}{\,\mu{\rm m}}
\newcommand{\cm}{\,{\rm cm}}
\newcommand{\meter}{\,{\rm m}}
\newcommand{\To}{\rightarrow}
\newcommand{\Ell}{l}
\newcommand{\Ds}{{\rm D_s}}
\newcommand{\Dsm}{{\rm D_s^-}}
\newcommand{\Dp}{{\rm D^+}}
\newcommand{\Bz}{{\rm B}^0}
\newcommand{\Bzbar}{\bar{\rm B}^0}
\newcommand{\Bs}{{\rm B^0_s}}
\newcommand{\Bsb}{\bar{\rm B}^0_{\rm s}}
\newcommand{\Kzbar}{\bar{\rm K}^0}
\newcommand{\Kp}{{\rm K^+}}
\newcommand{\Km}{{\rm K^-}}
\newcommand{\KKpi}{{\rm KK}\pi}
\newcommand{\fB}{f_{\rm B}}
\newcommand{\fDs}{f_{\Ds}}
\newcommand{\fDp}{f_{\rm D}}
\newcommand{\vcs}{V_{\rm cs}}
\newcommand{\vcd}{V_{\rm cd}}
\newcommand{\Rc}{R_{\rm c}}
\newcommand{\Rb}{R_{\rm b}}
\newcommand{\Z}{{\rm Z}}
\newcommand{\Uc}{{\rm U_c}}
\newcommand{\Ub}{{\rm U_b}}
\newcommand{\Uargs}{\Uci,\Ubi}
\newcommand{\Uci}{{\rm U}_{\rm c}^{(i)}}
\newcommand{\Ubi}{{\rm U}_{\rm b}^{(i)}}
\newcommand{\ee}{\rm e^+e^-}
\newcommand{\mumu}{\mu^+\mu^-}

\newcommand{\qqbar}{\rm q\bar{\rm q}}
\newcommand{\ccbar}{{\rm c\bar{\rm c}}}
\newcommand{\bbbar}{{\rm b\bar{\rm b}}}
\newcommand{\uds}{{\rm uds}}
\newcommand{\munu}{\mu\nu}
\newcommand{\taunu}{\tau\nu}
\newcommand{\Mmunu}{M_{\munu}}
\newcommand{\Dsenu}{\Ds\To {\rm e}\nu}
\newcommand{\Dsmunu}{\Ds\To\mu\nu}
\newcommand{\Dstaunu}{\Ds\To\tau\nu}
\newcommand{\Dslnu}{\Ds\To\Ell\nu}
\newcommand{\Dpmunu}{\Dp\To\mu\nu}
\newcommand{\Dptaunu}{\Dp\To\tau\nu}
\newcommand{\Dplnu}{\Dp\To\Ell\nu}
\newcommand{\tautoe}{\tau\To {\rm e}\nu\bar{\nu}}
\newcommand{\tautomu}{\tau\To\mu\nu\bar{\nu}}
\newcommand{\tautol}{\tau\To\Ell\nu\bar{\nu}}
\newcommand{\mununu}{\mu\nu\bar{\nu}}
\newcommand{\thethr}{\theta_{\rm thrust}}
\newcommand{\Dsphipi}{\Ds\To\phi\pi}
\newcommand{\gequ}{\,\protect\raisebox{-0.35ex}%
  {$\stackrel{\scriptstyle >}{\scriptstyle\sim}$}\,}
\newcommand{\hs}{\rule{3.5mm}{0pt}}
\newcommand{\hS}{\rule{6.0mm}{0pt}}
\begin{document}
{\setlength{\parskip}{4pt plus 1pt}%
\flushbottom
%-----------------------------------------------------------------------
\begin{titlepage}
\centerline{\large EUROPEAN ORGANIZATION FOR NUCLEAR RESEARCH}
\bigskip
\begin{flushright}
CERN--EP/2002--001\\
8 January 2002\\
\end{flushright}
\vfill
\begin{center}
{\LARGE\bf\boldmath Leptonic decays of the $\Ds$ meson}
\end{center}
\bigskip
\begin{center}
\renewcommand{\thefootnote}{\fnsymbol{footnote}}
\setcounter{footnote}{2}%
{\Large The ALEPH Collaboration}%
\footnote{See the following pages for the list of authors.}
\end{center}
\renewcommand{\thefootnote}{\arabic{footnote}}
\setcounter{footnote}{0}
\vfill
\begin{center}{\large Abstract}\end{center}
\begin{quote}
The purely leptonic decays $\Dstaunu$ and $\Dsmunu$ are studied
in a sample of four million hadronic $\Z$ decays collected 
with the ALEPH detector at the LEP $\ee$ collider
from 1991 to 1995.
The branching fractions are extracted from a combination
of two analyses,
one optimized to select $\Dstaunu$ decays
with $\tautoe$ or $\mununu$,
and the other optimized for $\Dsmunu$ decays.
The results are used to evaluate the $\Ds$ decay constant,
within the Standard Model:
$\fDs = [285 \pm 19({\rm stat}) \pm 40 ({\rm syst})] \mev$.
\end{quote}
\vfill

\begin{center}
Submitted to Physics Letters B
\end{center}

\bigskip\bigskip
\end{titlepage}
}%
% ----------------------------------------------------------------------
%------------------------------------------------------------------------
% authors12pt.tex
%for LEP I papers only
%-----------------------------------------------------------------------
\pagestyle{empty}
\newpage
\small
%
% remember the old settings
\newlength{\saveparskip}
\newlength{\savetextheight}
\newlength{\savetopmargin}
\newlength{\savetextwidth}
\newlength{\saveoddsidemargin}
\newlength{\savetopsep}
\newlength{\savevoffset}
\newlength{\savehoffset}
\setlength{\saveparskip}{\parskip}
\setlength{\savetextheight}{\textheight}
\setlength{\savetopmargin}{\topmargin}
\setlength{\savetextwidth}{\textwidth}
\setlength{\saveoddsidemargin}{\oddsidemargin}
\setlength{\savetopsep}{\topsep}
\setlength{\savevoffset}{\voffset}
\setlength{\savehoffset}{\hoffset}
%
% text dimensions for the author list
%
\setlength{\parskip}{0.0cm}
\setlength{\textheight}{25.0cm}
\setlength{\topmargin}{-1.5cm}
\setlength{\textwidth}{16 cm}
\setlength{\oddsidemargin}{-0.0cm}
\setlength{\topsep}{1mm}
\setlength{\voffset}{1mm}
\setlength{\hoffset}{0mm}
\pretolerance=10000
%%%%%%%%%\begin{document}
%\centerline{EUROPEAN ORGANIZATION FOR NUCLEAR RESEARCH}
%\centerline{EUROPEAN LABORATORY FOR PARTICLE PHYSICS (CERN)}
%\vspace{1cm}
%\begin{flushright}CERN-EP-2000-   \\
%17 December 2001 - last update
%\end{flushright}
\centerline{\large\bf The ALEPH Collaboration}
\footnotesize
\vspace{0.5cm}
{\raggedbottom
\begin{sloppypar}
\samepage\noindent
A.~Heister,
S.~Schael
\nopagebreak
\begin{center}
\parbox{15.5cm}{\sl\samepage
Physikalisches Institut das RWTH-Aachen, D-52056 Aachen, Germany}
\end{center}\end{sloppypar}
\vspace{2mm}
\begin{sloppypar}
\noindent
R.~Barate,
I.~De~Bonis,
D.~Decamp,
C.~Goy,
\mbox{J.-P.~Lees},
E.~Merle,
\mbox{M.-N.~Minard},
B.~Pietrzyk
\nopagebreak
\begin{center}
\parbox{15.5cm}{\sl\samepage
Laboratoire de Physique des Particules (LAPP), IN$^{2}$P$^{3}$-CNRS,
F-74019 Annecy-le-Vieux Cedex, France}
\end{center}\end{sloppypar}
\vspace{2mm}
\begin{sloppypar}
\noindent
G.~Boix,
S.~Bravo,
M.P.~Casado,
M.~Chmeissani,
J.M.~Crespo,
E.~Fernandez,
\mbox{M.~Fernandez-Bosman},
Ll.~Garrido,$^{15}$
E.~Graug\'{e}s,
M.~Martinez,
G.~Merino,
R.~Miquel,$^{27}$
Ll.M.~Mir,$^{27}$
A.~Pacheco,
H.~Ruiz
\nopagebreak
\begin{center}
\parbox{15.5cm}{\sl\samepage
Institut de F\'{i}sica d'Altes Energies, Universitat Aut\`{o}noma
de Barcelona, E-08193 Bellaterra (Barcelona), Spain$^{7}$}
\end{center}\end{sloppypar}
\vspace{2mm}
\begin{sloppypar}
\noindent
A.~Colaleo,
D.~Creanza,
M.~de~Palma,
G.~Iaselli,
G.~Maggi,
M.~Maggi,
S.~Nuzzo,
A.~Ranieri,
G.~Raso,$^{23}$
F.~Ruggieri,
G.~Selvaggi,
L.~Silvestris,
P.~Tempesta,
A.~Tricomi,$^{3}$
G.~Zito
\nopagebreak
\begin{center}
\parbox{15.5cm}{\sl\samepage
Dipartimento di Fisica, INFN Sezione di Bari, I-70126
Bari, Italy}
\end{center}\end{sloppypar}
\vspace{2mm}
\begin{sloppypar}
\noindent
X.~Huang,
J.~Lin,
Q. Ouyang,
T.~Wang,
Y.~Xie,
R.~Xu,
S.~Xue,
J.~Zhang,
L.~Zhang,
W.~Zhao
\nopagebreak
\begin{center}
\parbox{15.5cm}{\sl\samepage
Institute of High Energy Physics, Academia Sinica, Beijing, The People's
Republic of China$^{8}$}
\end{center}\end{sloppypar}
\vspace{2mm}
\begin{sloppypar}
\noindent
D.~Abbaneo,
P.~Azzurri,
O.~Buchm\"uller,$^{25}$
M.~Cattaneo,
F.~Cerutti,
B.~Clerbaux,
H.~Drevermann,
R.W.~Forty,
M.~Frank,
F.~Gianotti,
T.C.~Greening,$^{29}$
J.B.~Hansen,
J.~Harvey,
D.E.~Hutchcroft,
P.~Janot,
B.~Jost,
M.~Kado,$^{27}$
P.~Mato,
A.~Moutoussi,
F.~Ranjard,
L.~Rolandi,
D.~Schlatter,
O.~Schneider,$^{2}$
G.~Sguazzoni,
W.~Tejessy,
F.~Teubert,
A.~Valassi,
I.~Videau,
J.~Ward
\nopagebreak
\begin{center}
\parbox{15.5cm}{\sl\samepage
European Laboratory for Particle Physics (CERN), CH-1211 Geneva 23,
Switzerland}
\end{center}\end{sloppypar}
\vspace{2mm}
\begin{sloppypar}
\noindent
F.~Badaud,
A.~Falvard,$^{22}$
P.~Gay,
P.~Henrard,
J.~Jousset,
B.~Michel,
S.~Monteil,
\mbox{J-C.~Montret},
D.~Pallin,
P.~Perret
\nopagebreak
\begin{center}
\parbox{15.5cm}{\sl\samepage
Laboratoire de Physique Corpusculaire, Universit\'e Blaise Pascal,
IN$^{2}$P$^{3}$-CNRS, Clermont-Ferrand, F-63177 Aubi\`{e}re, France}
\end{center}\end{sloppypar}
\vspace{2mm}
\begin{sloppypar}
\noindent
J.D.~Hansen,
J.R.~Hansen,
P.H.~Hansen,
B.S.~Nilsson,
A.~W\"a\"an\"anen
\begin{center}
\parbox{15.5cm}{\sl\samepage
Niels Bohr Institute, DK-2100 Copenhagen, Denmark$^{9}$}
\end{center}\end{sloppypar}
\vspace{2mm}
\begin{sloppypar}
\noindent
A.~Kyriakis,
C.~Markou,
E.~Simopoulou,
A.~Vayaki,
K.~Zachariadou
\nopagebreak
\begin{center}
\parbox{15.5cm}{\sl\samepage
Nuclear Research Center Demokritos (NRCD), GR-15310 Attiki, Greece}
\end{center}\end{sloppypar}
\vspace{2mm}
\begin{sloppypar}
\noindent
A.~Blondel,$^{12}$
G.~Bonneaud,
\mbox{J.-C.~Brient},
A.~Roug\'{e},
M.~Rumpf,
M.~Swynghedauw,
M.~Verderi,
\linebreak
H.~Videau
\nopagebreak
\begin{center}
\parbox{15.5cm}{\sl\samepage
Laboratoire de Physique Nucl\'eaire et des Hautes Energies, Ecole
Polytechnique, IN$^{2}$P$^{3}$-CNRS, \mbox{F-91128} Palaiseau Cedex, France}
\end{center}\end{sloppypar}
\vspace{2mm}
%\pagebreak 
\begin{sloppypar}
\noindent
V.~Ciulli,
E.~Focardi,
G.~Parrini
\nopagebreak
\begin{center}
\parbox{15.5cm}{\sl\samepage
Dipartimento di Fisica, Universit\`a di Firenze, INFN Sezione di Firenze,
I-50125 Firenze, Italy}
\end{center}\end{sloppypar}
\vspace{2mm}
%\pagebreak
\begin{sloppypar}
\noindent
A.~Antonelli,
M.~Antonelli,
G.~Bencivenni,
G.~Bologna,$^{4}$
F.~Bossi,
P.~Campana,
G.~Capon,
V.~Chiarella,
P.~Laurelli,
G.~Mannocchi,$^{5}$
F.~Murtas,
G.P.~Murtas,
L.~Passalacqua,
\mbox{M.~Pepe-Altarelli},$^{24}$
P.~Spagnolo
\nopagebreak
\begin{center}
\parbox{15.5cm}{\sl\samepage
Laboratori Nazionali dell'INFN (LNF-INFN), I-00044 Frascati, Italy}
\end{center}\end{sloppypar}
\vspace{2mm}
%\pagebreak
\begin{sloppypar}
\noindent
A.~Halley,
J.G.~Lynch,
P.~Negus,
V.~O'Shea,
C.~Raine,$^{4}$
A.S.~Thompson
\nopagebreak
\begin{center}
\parbox{15.5cm}{\sl\samepage
Department of Physics and Astronomy, University of Glasgow, Glasgow G12
8QQ,United Kingdom$^{10}$}
\end{center}\end{sloppypar}
%\pagebreak
\vspace{2mm}
\begin{sloppypar}
\noindent
S.~Wasserbaech
\nopagebreak
\begin{center}
\parbox{15.5cm}{\sl\samepage
Department of Physics, Haverford College, Haverford, PA 19041-1392, U.S.A.}
\end{center}\end{sloppypar}
\vspace{2mm}
\pagebreak
\begin{sloppypar}
\noindent
R.~Cavanaugh,
S.~Dhamotharan,
C.~Geweniger,
P.~Hanke,
G.~Hansper,
V.~Hepp,
E.E.~Kluge,
A.~Putzer,
J.~Sommer,
H.~Stenzel,
K.~Tittel,
S.~Werner,$^{19}$
M.~Wunsch$^{19}$
\nopagebreak
\begin{center}
\parbox{15.5cm}{\sl\samepage
Kirchhoff-Institut f\"ur Physik, Universit\"at Heidelberg, D-69120
Heidelberg, Germany$^{16}$}
\end{center}\end{sloppypar}
\vspace{2mm}
\begin{sloppypar}
\noindent
R.~Beuselinck,
D.M.~Binnie,
W.~Cameron,
P.J.~Dornan,
M.~Girone,$^{1}$
N.~Marinelli,
J.K.~Sedgbeer,
J.C.~Thompson$^{14}$
\nopagebreak
\begin{center}
\parbox{15.5cm}{\sl\samepage
Department of Physics, Imperial College, London SW7 2BZ,
United Kingdom$^{10}$}
\end{center}\end{sloppypar}
\vspace{2mm}
\begin{sloppypar}
\noindent
V.M.~Ghete,
P.~Girtler,
E.~Kneringer,
D.~Kuhn,
G.~Rudolph
\nopagebreak
\begin{center}
\parbox{15.5cm}{\sl\samepage
Institut f\"ur Experimentalphysik, Universit\"at Innsbruck, A-6020
Innsbruck, Austria$^{18}$}
\end{center}\end{sloppypar}
\vspace{2mm}
\begin{sloppypar}
\noindent
E.~Bouhova-Thacker,
C.K.~Bowdery,
A.J.~Finch,
F.~Foster,
G.~Hughes,
R.W.L.~Jones,
M.R.~Pearson,
N.A.~Robertson
\nopagebreak
\begin{center}
\parbox{15.5cm}{\sl\samepage
Department of Physics, University of Lancaster, Lancaster LA1 4YB,
United Kingdom$^{10}$}
\end{center}\end{sloppypar}
\vspace{2mm}
\begin{sloppypar}
\noindent
K.~Jakobs,
K.~Kleinknecht,
G.~Quast,$^{6}$
B.~Renk,
\mbox{H.-G.~Sander},
H.~Wachsmuth,
C.~Zeitnitz
\nopagebreak
\begin{center}
\parbox{15.5cm}{\sl\samepage
Institut f\"ur Physik, Universit\"at Mainz, D-55099 Mainz, Germany$^{16}$}
\end{center}\end{sloppypar}
\vspace{2mm}
\begin{sloppypar}
\noindent
A.~Bonissent,
J.~Carr,
P.~Coyle,
O.~Leroy,
P.~Payre,
D.~Rousseau,
M.~Talby
\nopagebreak
\begin{center}
\parbox{15.5cm}{\sl\samepage
Centre de Physique des Particules, Universit\'e de la M\'editerran\'ee,
IN$^{2}$P$^{3}$-CNRS, F-13288 Marseille, France}
\end{center}\end{sloppypar}
\vspace{2mm}
%\pagebreak %jw
\begin{sloppypar}
\noindent
F.~Ragusa
\nopagebreak
\begin{center}
\parbox{15.5cm}{\sl\samepage
Dipartimento di Fisica, Universit\`a di Milano e INFN Sezione di Milano,
I-20133 Milano, Italy}
\end{center}\end{sloppypar}
\vspace{2mm}
\begin{sloppypar}
\noindent
A.~David,
H.~Dietl,
G.~Ganis,$^{26}$
K.~H\"uttmann,
G.~L\"utjens,
C.~Mannert,
W.~M\"anner,
\mbox{H.-G.~Moser},
R.~Settles,
W.~Wiedenmann,
G.~Wolf
\nopagebreak
\begin{center}
\parbox{15.5cm}{\sl\samepage
Max-Planck-Institut f\"ur Physik, Werner-Heisenberg-Institut,
D-80805 M\"unchen, Germany\footnotemark[16]}
\end{center}\end{sloppypar}
\vspace{2mm}
%\pagebreak
\begin{sloppypar}
\noindent
J.~Boucrot,
O.~Callot,
M.~Davier,
L.~Duflot,
\mbox{J.-F.~Grivaz},
Ph.~Heusse,
A.~Jacholkowska,
J.~Lefran\c{c}ois,
\mbox{J.-J.~Veillet},
C.~Yuan
\nopagebreak
\begin{center}
\parbox{15.5cm}{\sl\samepage
Laboratoire de l'Acc\'el\'erateur Lin\'eaire, Universit\'e de Paris-Sud,
IN$^{2}$P$^{3}$-CNRS, F-91898 Orsay Cedex, France}
\end{center}\end{sloppypar}
\vspace{2mm}
\begin{sloppypar}
\noindent
%\samepage
G.~Bagliesi,
T.~Boccali,
L.~Fo\`{a},
A.~Giammanco,
A.~Giassi,
F.~Ligabue,
A.~Messineo,
F.~Palla,
G.~Sanguinetti,
A.~Sciab\`a,
R.~Tenchini,$^{1}$
A.~Venturi,$^{1}$
P.G.~Verdini
\samepage
\begin{center}
\parbox{15.5cm}{\sl\samepage
Dipartimento di Fisica dell'Universit\`a, INFN Sezione di Pisa,
e Scuola Normale Superiore, I-56010 Pisa, Italy}
\end{center}\end{sloppypar}
\vspace{2mm}
\begin{sloppypar}
\noindent
G.A.~Blair,
G.~Cowan,
M.G.~Green,
T.~Medcalf,
A.~Misiejuk,
J.A.~Strong,
\mbox{P.~Teixeira-Dias},
\mbox{J.H.~von~Wimmersperg-Toeller}
\nopagebreak
\begin{center}
\parbox{15.5cm}{\sl\samepage
Department of Physics, Royal Holloway \& Bedford New College,
University of London, Egham, Surrey TW20 OEX, United Kingdom$^{10}$}
\end{center}\end{sloppypar}
\vspace{2mm}
\begin{sloppypar}
\noindent
R.W.~Clifft,
T.R.~Edgecock,
P.R.~Norton,
I.R.~Tomalin
\nopagebreak
\begin{center}
\parbox{15.5cm}{\sl\samepage
Particle Physics Dept., Rutherford Appleton Laboratory,
Chilton, Didcot, Oxon OX11 OQX, United Kingdom$^{10}$}
\end{center}\end{sloppypar}
\vspace{2mm}
%\pagebreak
\begin{sloppypar}
\noindent
\mbox{B.~Bloch-Devaux},
P.~Colas,
S.~Emery,
W.~Kozanecki,
E.~Lan\c{c}on,
\mbox{M.-C.~Lemaire},
E.~Locci,
P.~Perez,
J.~Rander,
\mbox{J.-F.~Renardy},
A.~Roussarie,
\mbox{J.-P.~Schuller},
J.~Schwindling,
A.~Trabelsi,$^{21}$
B.~Vallage
\nopagebreak
\begin{center}
\parbox{15.5cm}{\sl\samepage
CEA, DAPNIA/Service de Physique des Particules,
CE-Saclay, F-91191 Gif-sur-Yvette Cedex, France$^{17}$}
\end{center}\end{sloppypar}
%\pagebreak
\vspace{2mm}
\begin{sloppypar}
\noindent
N.~Konstantinidis,
A.M.~Litke,
G.~Taylor
\nopagebreak
\begin{center}
\parbox{15.5cm}{\sl\samepage
Institute for Particle Physics, University of California at
Santa Cruz, Santa Cruz, CA 95064, USA$^{13}$}
\end{center}\end{sloppypar}
\vspace{2mm}
\begin{sloppypar}
\noindent
C.N.~Booth,
S.~Cartwright,
F.~Combley,$^{4}$
M.~Lehto,
L.F.~Thompson
\nopagebreak
\begin{center}
\parbox{15.5cm}{\sl\samepage
Department of Physics, University of Sheffield, Sheffield S3 7RH,
United Kingdom$^{10}$}
\end{center}\end{sloppypar}
\vspace{2mm}
\pagebreak
\begin{sloppypar}
\noindent
K.~Affholderbach,$^{28}$
A.~B\"ohrer,
S.~Brandt,
C.~Grupen,
A.~Ngac,
G.~Prange,
U.~Sieler
\nopagebreak
\begin{center}
\parbox{15.5cm}{\sl\samepage
Fachbereich Physik, Universit\"at Siegen, D-57068 Siegen,
 Germany$^{16}$}
\end{center}\end{sloppypar}
\vspace{2mm}
\begin{sloppypar}
\noindent
G.~Giannini
\nopagebreak
\begin{center}
\parbox{15.5cm}{\sl\samepage
Dipartimento di Fisica, Universit\`a di Trieste e INFN Sezione di Trieste,
I-34127 Trieste, Italy}
\end{center}\end{sloppypar}
\vspace{2mm}
\begin{sloppypar}
\noindent
H.~He,
J.~Putz,
J.~Rothberg
\nopagebreak
\begin{center}
\parbox{15.5cm}{\sl\samepage
Experimental Elementary Particle Physics, University of Washington, Seattle, 
WA 98195 U.S.A.}
\end{center}\end{sloppypar}
\vspace{2mm}
\begin{sloppypar}
\noindent
S.R.~Armstrong,
K.~Berkelman,
K.~Cranmer,
D.P.S.~Ferguson,
Y.~Gao,$^{20}$
S.~Gonz\'{a}lez,
O.J.~Hayes,
H.~Hu,
S.~Jin,
J.~Kile,
P.A.~McNamara III,
J.~Nielsen,
Y.B.~Pan,
\mbox{J.H.~von~Wimmersperg-Toeller},
W.~Wiedenmann,
J.~Wu,
Sau~Lan~Wu,
X.~Wu,
G.~Zobernig
\nopagebreak
\begin{center}
\parbox{15.5cm}{\sl\samepage
Department of Physics, University of Wisconsin, Madison, WI 53706,
USA$^{11}$}
\end{center}\end{sloppypar}
\vspace{2mm}
\begin{sloppypar}
\noindent
G.~Dissertori
\nopagebreak
\begin{center}
\parbox{15.5cm}{\sl\samepage
Institute for Particle Physics, ETH H\"onggerberg, 8093 Z\"urich, Switzerland.}
\end{center}\end{sloppypar}
}
\footnotetext[1]{Also at CERN, 1211 Geneva 23, Switzerland.}
\footnotetext[2]{Now at Universit\'e de Lausanne, 1015 Lausanne, Switzerland.}
\footnotetext[3]{Also at Dipartimento di Fisica di Catania and INFN Sezione di
 Catania, 95129 Catania, Italy.}
\footnotetext[4]{Deceased.}
\footnotetext[5]{Also Istituto di Cosmo-Geofisica del C.N.R., Torino,
Italy.}
\footnotetext[6]{Now at Institut f\"ur Experimentelle Kernphysik, Universit\"at Karlsruhe, 76128 Karlsruhe, Germany.}
\footnotetext[7]{Supported by CICYT, Spain.}
\footnotetext[8]{Supported by the National Science Foundation of China.}
\footnotetext[9]{Supported by the Danish Natural Science Research Council.}
\footnotetext[10]{Supported by the UK Particle Physics and Astronomy Research
Council.}
\footnotetext[11]{Supported by the US Department of Energy, grant
DE-FG0295-ER40896.}
\footnotetext[12]{Now at Departement de Physique Corpusculaire, Universit\'e de
Gen\`eve, 1211 Gen\`eve 4, Switzerland.}
\footnotetext[13]{Supported by the US Department of Energy,
grant DE-FG03-92ER40689.}
\footnotetext[14]{Also at Rutherford Appleton Laboratory, Chilton, Didcot, UK.}
\footnotetext[15]{Permanent address: Universitat de Barcelona, 08208 Barcelona,
Spain.}
\footnotetext[16]{Supported by the Bundesministerium f\"ur Bildung,
Wissenschaft, Forschung und Technologie, Germany.}
\footnotetext[17]{Supported by the Direction des Sciences de la
Mati\`ere, C.E.A.}
\footnotetext[18]{Supported by the Austrian Ministry for Science and Transport.}
\footnotetext[19]{Now at SAP AG, 69185 Walldorf, Germany.}
\footnotetext[20]{Also at Department of Physics, Tsinghua University, Beijing, The People's Republic of China.}
\footnotetext[21]{Now at D\'epartement de Physique, Facult\'e des Sciences de Tunis, 1060 Le Belv\'ed\`ere, Tunisia.}
\footnotetext[22]{Now at Groupe d'Astroparticules de Montpellier, Universit\'{e} de Montpellier II, 34095, Montpellier, France}
\footnotetext[23]{Also at Dipartimento di Fisica e Tecnologie Relative, Universit\`a di Palermo, Palermo, Italy.}
\footnotetext[24]{Now at CERN, 1211 Geneva 23, Switzerland.}
\footnotetext[25]{Now at SLAC, Stanford, CA 94309, U.S.A.}
\footnotetext[26]{Now at INFN Sezione di Roma II, Dipartimento di Fisica, Universit\'a di Roma Tor Vergata, 00133 Roma, Italy.} 
\footnotetext[27]{Now at LBNL, Berkeley, CA 94720, U.S.A.}
\footnotetext[28]{Now at Skyguide, Swissair Navigation Services, Geneva, Switzerland.}
\footnotetext[29]{Now at Honeywell, Phoenix AZ, U.S.A.}
% restore the previous settings
\setlength{\parskip}{\saveparskip}
\setlength{\textheight}{\savetextheight}
\setlength{\topmargin}{\savetopmargin}
\setlength{\textwidth}{\savetextwidth}
\setlength{\oddsidemargin}{\saveoddsidemargin}
\setlength{\topsep}{\savetopsep}
\setlength{\voffset}{\savevoffset}
\setlength{\hoffset}{\savehoffset}
%%%%%%%%%%%%%%%%%%%%%%%%%%%%%%%%%%%%%%%%%
\normalsize
\newpage
\pagestyle{plain}
\setcounter{page}{1}
%-----------------------------------------------------------------------
\setlength{\parskip}{4pt plus 1pt}%
\raggedbottom%
\section{Introduction}
\label{s:intro}%
The leptonic decays of the $\Ds$ meson are interesting because
they open a window onto the strong interactions of the constituent
quarks of the $\Ds$.
The absence of strong interactions among the final state particles
makes the interpretation of the measurements particularly
straightforward.
The information gained is useful in understanding other processes
involving heavy pseudoscalar mesons.

The decay $\Dslnu$ is the second-generation analogue of
charged pion decay, $\pi^+\To\Ell\nu$.
The $\Dslnu$ decay proceeds via the
Cabibbo allowed annihilation of the
c and $\bar{\rm s}$ quarks in the $\Ds$;
the rate depends on the ${\rm c}\bar{\rm s}$ wavefunction
at zero separation\rlap.\footnote{%
Charge-conjugate states are implied throughout this paper.}
The annihilation amplitude is characterized by
the $\Ds$ decay constant $\fDs$, defined as
\begin{equation}
\label{eq:fds}
i \fDs p^{(\Ds)}_{\mu} =
\langle \Ds | \bar{\rm c} \gamma_{\mu} \gamma_{5} {\rm s} | 0 \rangle\ .
\end{equation}
In the Standard Model the leptonic branching fraction is then given by
\begin{equation}
\label{eq:lepbr}
B(\Dslnu) =
\frac{G_{\rm F}^2}{8\pi} \,
\tau_{\Ds} \,
\fDs^2 \,
|\vcs|^2 \,
m_{\Ds} \,
m_{l}^2
\left(1 - \frac{m_{l}^2}{m_{\Ds}^2}\right)^2,
\end{equation}
where
$\tau_{\Ds}$ is the mean lifetime of the $\Ds$,
$\vcs$ is the relevant CKM matrix element,
$m_{\Ds}$ is the mass of the $\Ds$, and
$m_l$ is the mass of the lepton.

It is worthwhile to measure $\fDs$
because it characterizes the structure of the $\Ds$ meson and
can be calculated in various theoretical models.
Lattice QCD is now generally considered to be the most
reliable method for calculating the pseudoscalar meson decay constants;
recent results yield a prediction of
$\fDs = 255 \pm 30 \mev$~\cite{bernard}.
An important application of the lattice QCD calculations
of decay constants
is in the evaluation of the third-generation 
Cabibbo-Kobayashi-Maskawa matrix elements.
Constraints on these elements are obtained from
measurements of $\Bz$-$\Bzbar$ mixing
and rely on theoretical estimates of $\fB$,
which is experimentally inaccessible at present.
A measurement of $\fDs$ therefore serves as a useful
check of the theoretical methods.

The relative branching fractions for
$\Dsenu$, $\munu$, and $\taunu$
are known precisely in the Standard Model.
From Eq.~\ref{eq:lepbr},
the ratios of branching fractions are 
$B(\Dsenu) /B(\Dstaunu) = 2.5 \times 10^{-6}$ and
$B(\Dsmunu)/B(\Dstaunu) = 0.103$.
The helicity suppression in these decays
leads to an extremely small branching fraction
for the electron mode.
An investigation of the muon and tau modes is
presented in this paper.

The helicity suppression is lifted if a photon is radiated
from one of the quarks, giving $\Ds\To\gamma\Ell\nu$.
The analysis of Ref.~\cite{burdman} leads to an estimate
of the ratio
$r = B(\Ds\To\gamma\mu\nu)/B(\Dsmunu) \cong 0.014$ to $0.11$
with
$B(\Ds\To\gamma{\rm e}\nu) \cong B(\Ds\To\gamma\mu\nu)$.
The prediction in~\cite{atwood} is smaller,
$r \cong 0.0059$.
In either case, the effect of $\Ds\To\gamma\Ell\nu$ decays
on the present investigation would be small or negligible,
and no correction is made for these decays.

The data sample and detector
are briefly described in Section~\ref{s:detdata}.
Two analyses are then presented,
one optimized to select $\Dstaunu$ decays (Section~\ref{s:taunu})
and one optimized to select $\Dsmunu$ decays (Section~\ref{s:munu}).
The systematic errors are discussed in Section~\ref{s:syst}, and
the combined results and conclusions are given in Section~\ref{s:conc}.

% ----------------------------------------------------------------------
\section{Data sample and apparatus}
\label{s:detdata}%
The analyses are based on a data sample collected 
with the ALEPH detector
at the LEP $\ee$ collider
in 1991 through 1995,
at or near the peak of the $\Z$ resonance.
The sample corresponds to $3.97\times 10^{6}$ produced
$\ee\To\qqbar$ events.

Detailed descriptions of the ALEPH detector and its performance may
be found in~\cite{detect,perf}.
%{\catcode`@=11[\hbox{\b@detect}--\hbox{\b@perf}]}.
The tracking system consists of
a high-resolution silicon strip vertex detector (VDET),
a cylindrical drift chamber (the inner tracking chamber or ITC),
and a large time projection chamber (TPC).
The VDET comprises two layers
of double-sided silicon strip detectors
at average radii of $6.3$ and $10.8\cm$.
The spatial resolution for the $r$-$\phi$ and $z$ projections
(transverse to and along the beam axis, respectively)
is $12 \mic$ at normal incidence.
The angular coverage is
$\left|\cos\theta\right| < 0.85$ for the inner layer and
$\left|\cos\theta\right| < 0.69$ for the outer layer.
The ITC has eight coaxial wire layers
at radii from 16 to $26\cm$.
The TPC provides up to 21 three-dimensional coordinates per track
at radii between 40 and $171\cm$,
as well as measurements of the specific ionization energy loss
($dE/dx$) of charged particles.
The tracking detectors are contained within
a superconducting solenoid,
which produces an axial magnetic field of $1.5\,{\rm T}$.
Charged tracks measured in the VDET-ITC-TPC system
are reconstructed with a momentum resolution of
$\Delta p/p = 
6\times 10^{-4} \, p_{t} \oplus 0.005$ 
($p_{t}$ in ${\rm GeV}/c$).
An impact parameter resolution of $22\mic$ in the $r$-$\phi$ plane
is achieved for muons from $\Z\rightarrow\mumu$ having at least
one VDET $r$-$\phi$ hit.
These performance figures reflect the improvements
obtained from the 1998 reprocessing of the 
ALEPH LEP1 data set
with enhanced event reconstruction algorithms~\cite{reproc}.

Surrounding the TPC is an electromagnetic calorimeter (ECAL),
a lead/wire-chamber sandwich operated in proportional mode.
The calorimeter is read out via projective towers
subtending typically $0.9^{\circ} \times 0.9^{\circ}$ 
which sum the deposited energy in three sections in depth.
Beyond the ECAL lies the solenoid,
followed by a hadron calorimeter (HCAL),
which uses the iron return yoke as absorber
and has an average depth of $1.50\meter$.
Hadronic showers are sampled by 23 planes of streamer tubes,
providing a digital hit pattern and
inducing an analog signal on pads arranged in projective towers.
The HCAL is used in combination with two layers of muon chambers
outside the magnet for $\mu$ identification.

The measurements of charged particle momenta and of 
energy depositions in the calorimeters,
combined with the identification of photons, electrons,
and muons,
are used to produce a list of charged and neutral
{\it energy flow particles\/} in each event~\cite{perf}.

% ----------------------------------------------------------------------
\section{\boldmath$\Dstaunu$ analysis}
\label{s:taunu}%
The first analysis~\cite{harry} is
designed to search for the decay chain
$\ee\To\ccbar$,
${\rm c}\To\Dstaunu$,
with $\tautoe$ or $\tautomu$.
The electron and muon channels are treated separately.
As the final state under study contains three neutrinos,
the signature of these decays is an identified lepton 
and large missing energy in one hemisphere of the event. 

Although the $\ee\To\bbbar$ event sample contains more $\Ds$ mesons
than the $\ccbar$ sample,
the $\Dslnu$ signal is more difficult to isolate in $\bbbar$ events
due to the softer $\Ds$ spectrum and
the large number of semileptonic b decays.
Cuts are therefore applied to remove $\bbbar$ events,
and the analyses are optimized to select $\Ds$ decays
in $\ccbar$ events.

\subsection{Event selection}
\label{s:taunu:sel}%
Hadronic $\Z$ decays are preselected
using charged tracks~\cite{Z}.
Backgrounds from two-photon interactions and
dilepton events are reduced to a negligible level
by means of additional cuts
on the numbers of reconstructed charged and neutral particles.
The event thrust axis is required to satisfy $|{ \cos\thethr }| < 0.8$
to select events within the acceptance of the vertex detector. 
Each event is then separated
into two hemispheres by the plane perpendicular to the thrust axis. 
The total energy of each hemisphere is calculated
and the hemisphere with larger missing energy is selected 
if it contains an identified lepton (e or $\mu$).
Electron identification is based on the shower shape 
in the electromagnetic calorimeter and 
the ionization in the time projection chamber;
muon identification makes use of the digital pattern information in the
hadron calorimeter and
hits in the muon chambers~\cite{qselep}.
If more than one lepton is present in the selected hemisphere,
the one with the highest momentum is considered.
The missing energy in the hemisphere is required to be 
greater than $5\gev$ to reduce the background.
The hemisphere invariant masses are taken into account
in the calculation of the expected hemisphere energies~\cite{bslife}.

To further reduce the background from $\bbbar$ events,
cuts are applied to a set of hemisphere variables
based on the pseudorapidities and impact parameters of the charged tracks.
The energy flow particles are clustered into jets
by means of the JADE algorithm with $y_{\rm cut} = 0.003$.
The pseudorapidity $\eta$ of a particle
is then defined as $\eta = -\ln\tan(\alpha/2)$, where 
$\alpha$ is the laboratory angle between the particle and 
the nearest jet axis.
The reconstructed particles in $\ccbar$ events
tend to have larger pseudorapidities than those
in $\bbbar$ events
because of the lower mass of c hadrons compared to b hadrons.
An existing ALEPH lifetime-based b tag algorithm~\cite{qipbtag} was 
modified to include a dependence on 
the pseudorapidity of the charged tracks.
The tracks in each hemisphere are divided into two groups,
one with $\eta \geq 5.1$ and one with $\eta < 5.1$.
For each group,
the confidence level for the hypothesis that all tracks
originate from the $\Z$ production point is then calculated.
The selected lepton track is not included in this calculation.
To reject $\bbbar$ events,
a cut is made on the confidence level for the low-$\eta$
group in each hemisphere.
A further rejection of $\ccbar$ background events\footnote{%
Throughout this paper, light quark-antiquark (denoted ``$\uds$''), 
$\ccbar$, and $\bbbar$ background
events are defined to be those events containing no
$\Dslnu$ or $\Dplnu$ decays.}
is obtained by cutting on the confidence level for the
high-$\eta$ group in the hemisphere containing the lepton;
with the lepton excluded,
signal events have only fragmentation tracks in the
lepton hemisphere
and can be distinguished from $\ccbar$ background events.

The momentum and energy of the $\Ds$ candidate are reconstructed 
from the observed charged and neutral particles in the event
by excluding the lepton $l$ and applying four-momentum conservation:
\begin{eqnarray*}
\vec{P}_{\Ds} &=& -\sum_{i\neq l} \vec{P_{i}} \\
E_{\Ds}       &=& \sqrt{s} - \sum_{i\neq l} E_{i} \ .
\end{eqnarray*}
This procedure is based on the assumption that no neutrinos are
produced in the hemisphere opposite the $\Ds$ candidate.
A kinematic fit is performed in which the energies of all reconstructed
particles (except the lepton) are varied such that the constraint
$[E_{\Ds}^{2} - P_{\Ds}^{2}c^2 ]^{1/2} = M_{\Ds}c^2 = 1.968\gev$ 
is satisfied. 
The energy resolutions for charged, neutral electromagnetic,
and neutral hadronic particles are parametrized from simulated events.
The kinematic fit improves the energy resolution of $\Ds$ candidates 
from $6.7\gev$ to $3.5\gev$. 
The background is further reduced by the requirement that
the fitted $\Ds$ energy be greater than $22.5\gev$.
The selection efficiency for
$\ccbar\To\Dstaunu$ events is 2.5\% (3.3\%)
in the electron (muon) channel
[including the factor of $B(\tautol)$].
This procedure selects 3956 and 6637 events in the electron
and muon channels, respectively.

\subsection{Linear discriminant analysis}
\begin{figure}[p]
\begin{center}
\mbox{\includegraphics[height=150mm, bb=0 0 525 525]{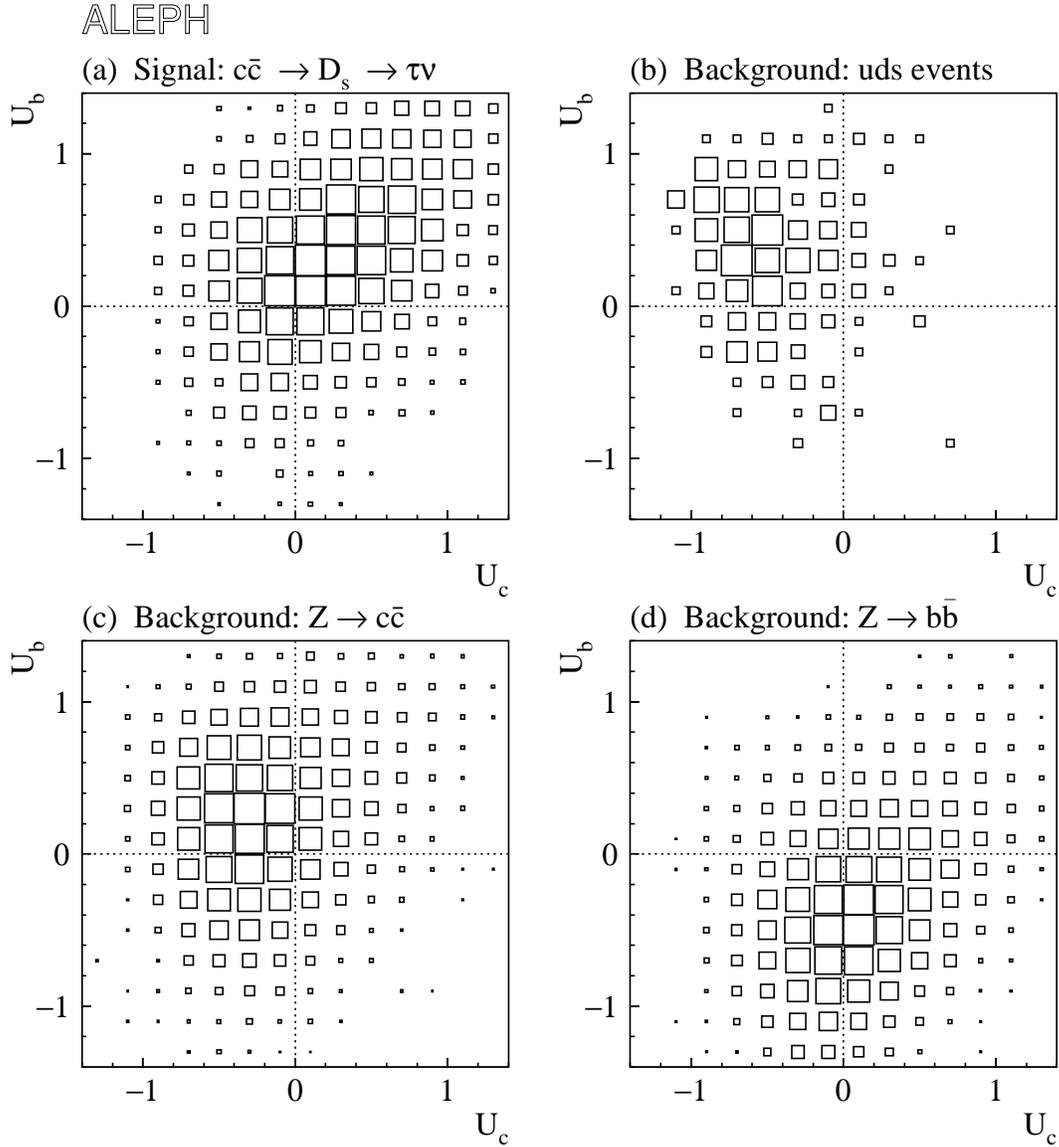}}
\end{center}
\caption{\label{f:utaue}
Monte Carlo $\Ub$~versus $\Uc$ distributions in the electron channel of the
$\taunu$ analysis, after all cuts:
(a) signal, $\ccbar\To\Dstaunu$;
(b) uds background;
(c) $\ccbar$ background;
(d) $\bbbar$ background.
The area of the square in each bin is proportional to the
number of entries.
The distributions are shown with arbitrary normalizations.}
\end{figure}

\begin{figure}[p]
\begin{center}
\mbox{\includegraphics[height=150mm, bb=0 0 525 525]{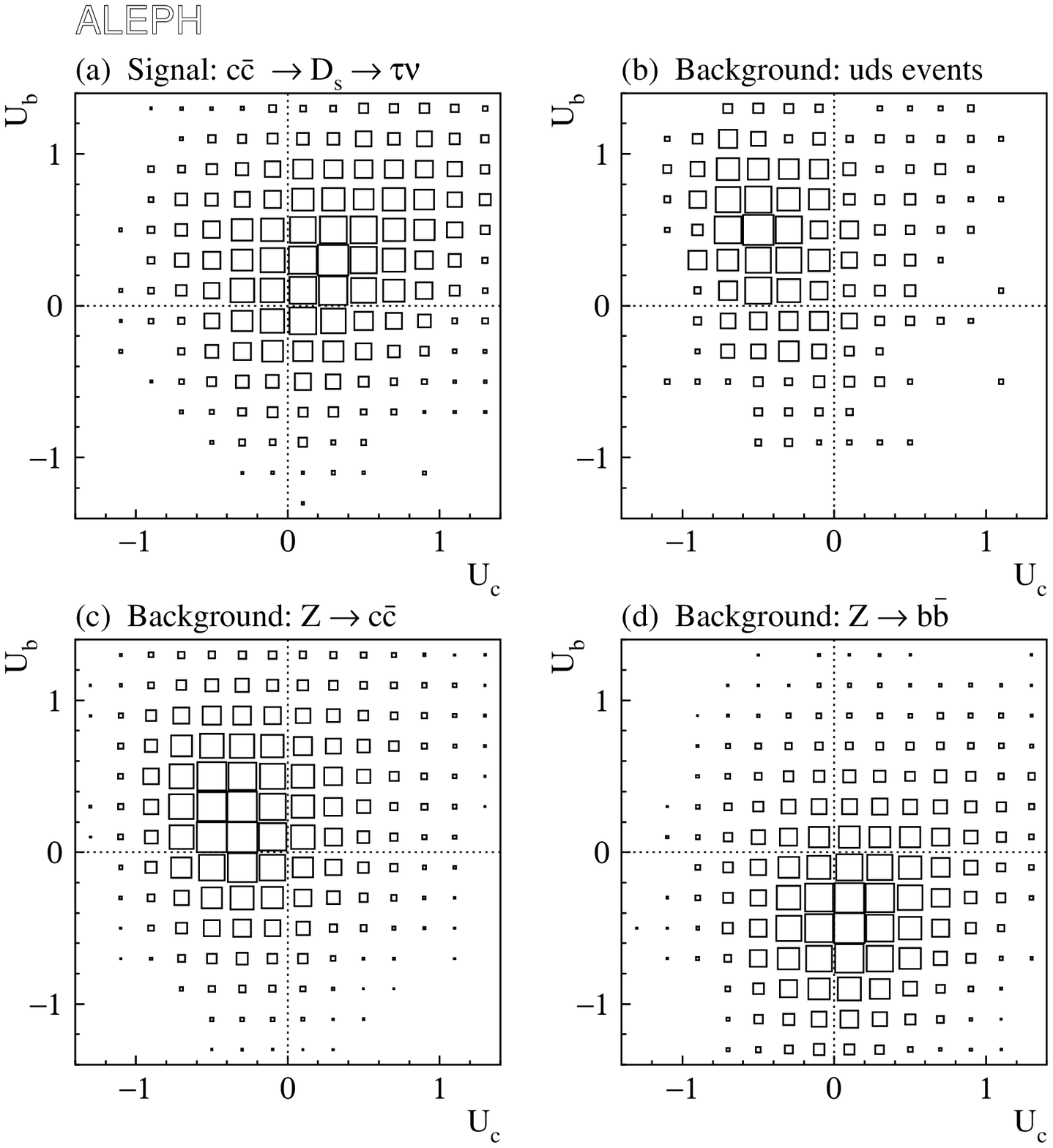}}
\end{center}
\caption{\label{f:utaum}
Monte Carlo $\Ub$~versus $\Uc$ distributions in the muon channel of the
$\taunu$ analysis, after all cuts:
(a) signal, $\ccbar\To\Dstaunu$;
(b) uds background;
(c) $\ccbar$ background;
(d) $\bbbar$ background.
The distributions are shown with arbitrary normalizations.}
\end{figure}

A linear discriminant analysis is performed on Monte Carlo events 
to search for an optimal linear combination of event variables so that 
maximum discrimination between signal and background is achieved.
Two discriminating variables, $\Uc$ and $\Ub$, are created to distinguish 
the $\ccbar\To\Dstaunu$ signal events from $\ccbar$ and $\bbbar$ 
backgrounds, respectively;
the backgrounds arise mainly from semileptonic c and b decays.
Nine variables are selected to form $\Uc$ and ten for $\Ub$.
The variables with the greatest discrimination power include
the fitted $\Ds$ momentum,
the angle between the lepton and the $\Ds$ direction 
in the $\Ds$ rest frame, 
several b tag variables, and 
the lepton transverse momentum with respect to the nearest jet axis.
The definitions of $\Uc$ and $\Ub$ are optimized independently
in the electron and muon channels.
The distributions of $\Ub$~versus $\Uc$ are shown in 
Figs.~\ref{f:utaue} and \ref{f:utaum}.

The branching fraction $B(\Dstaunu)$ is extracted from 
a maximum likelihood fit to the 
unbinned $\Ub$~versus $\Uc$ distribution in each channel. 
The two-dimensional fitting function 
consists of one signal and three background 
($\uds$, $\ccbar$, and $\bbbar$) components.
Although the analysis is optimized to select $\Dstaunu$ decays,
some $\Dsmunu$ decays are also accepted.
Because these two decay rates are both proportional to $\fDs^2$,
both types of decays are considered to be part of the signal
in the fit to the data.
Leptonic $\Dp$ decays also pass the selection and
are included in the signal.
The production of $\Ds$ and $\Dp$ mesons in
the remaining $\bbbar$ events must be taken into account as well.
The signal component of the fitting function
is therefore taken to be the combination of 
$\Ds$ and $\Dp$ decays to $\taunu$ and $\munu$
in $\ccbar$ and $\bbbar$ events.
The relative normalizations of these eight contributions
are fixed according to Eq.~\ref{eq:lepbr} and 
the measured charm production rates~\cite{lepewwg,ccount};
$\fDs/\fDp = 1.11\,{}^{+0.06}_{-0.05}$ 
from lattice QCD calculations~\cite{draper};
$\Rc$ and $\Rb$ from the Standard Model fit in~\cite{lepewwg};
$|{ \vcs }|$ and $|{ \vcd }|$ from~\cite{rpp};
the $\Ds$ and $\Dp$ lifetimes from~\cite{rpp}; 
and particle masses from~\cite{rpp}.
The normalizations are shown in Table~\ref{t:tausig}.
Leptonic $\Dp$ decays are Cabibbo suppressed and contribute
less than 8\% of the total signal.
Each component of the fitting function is parametrized as 
the sum of up to five two-dimensional correlated Gaussian functions
from a fit to simulated events.
\begin{table}[t]
\begin{center}
\caption{Relative normalizations of the contributions 
to the signal in the $\Dstaunu$ analysis.}{\label{t:tausig}}
\vspace{2mm}
\begin{tabular}{lrr}                                       \hline
         & \multicolumn{2}{c}{Fraction of signal (\%)}   \\ 
Source   & \multicolumn{1}{l}{e channel} & \multicolumn{1}{l}{$\mu$ channel} \\ \hline
$\ccbar\To\Dstaunu$   &  76.9\hs &  52.4\hs  \\
$\ccbar\To\Dsmunu$    &   0.5\hs &  26.4\hs  \\
$\bbbar\To\Dstaunu$   &  17.4\hs &  10.7\hs  \\
$\bbbar\To\Dsmunu$    &   1.2\hs &   3.1\hs  \\
$\ccbar\To\Dptaunu$   &   3.4\hs &   2.3\hs  \\
$\ccbar\To\Dpmunu$    &   0.1\hs &   4.4\hs  \\
$\bbbar\To\Dptaunu$   &   0.4\hs &   0.4\hs  \\
$\bbbar\To\Dpmunu$    &   0.2\hs &   0.4\hs  \\ \hline
Total                 & 100.0\hs & 100.0\hs  \\ \hline
\end{tabular}
\end{center}
\end{table}

A program based on {\tt JETSET} 7.4~\cite{jetset} 
and tuned to ALEPH data~\cite{tuning} was used to generate
the Monte Carlo sample of 15 million 
$\ee\To\qqbar$ background events
(including 3.7 million $\ccbar$ and 6.6 million $\bbbar$ events)
and $370\,000$ events containing $\Ds$ and $\Dp$ leptonic decays.
The polarization of the $\tau$ is taken into account in the simulation
of $({\rm pseudoscalar}) \To\tau\nu$ decays.

The free parameters in the fit to the data are the
numbers of events in the signal and the three background components.
The fitting procedure is tested with Monte Carlo events 
to be free of bias.

\subsection{Results}
\label{s:taunu:results}%
\begin{figure}[p]
\begin{center}
\mbox{%
\includegraphics[height=150mm]{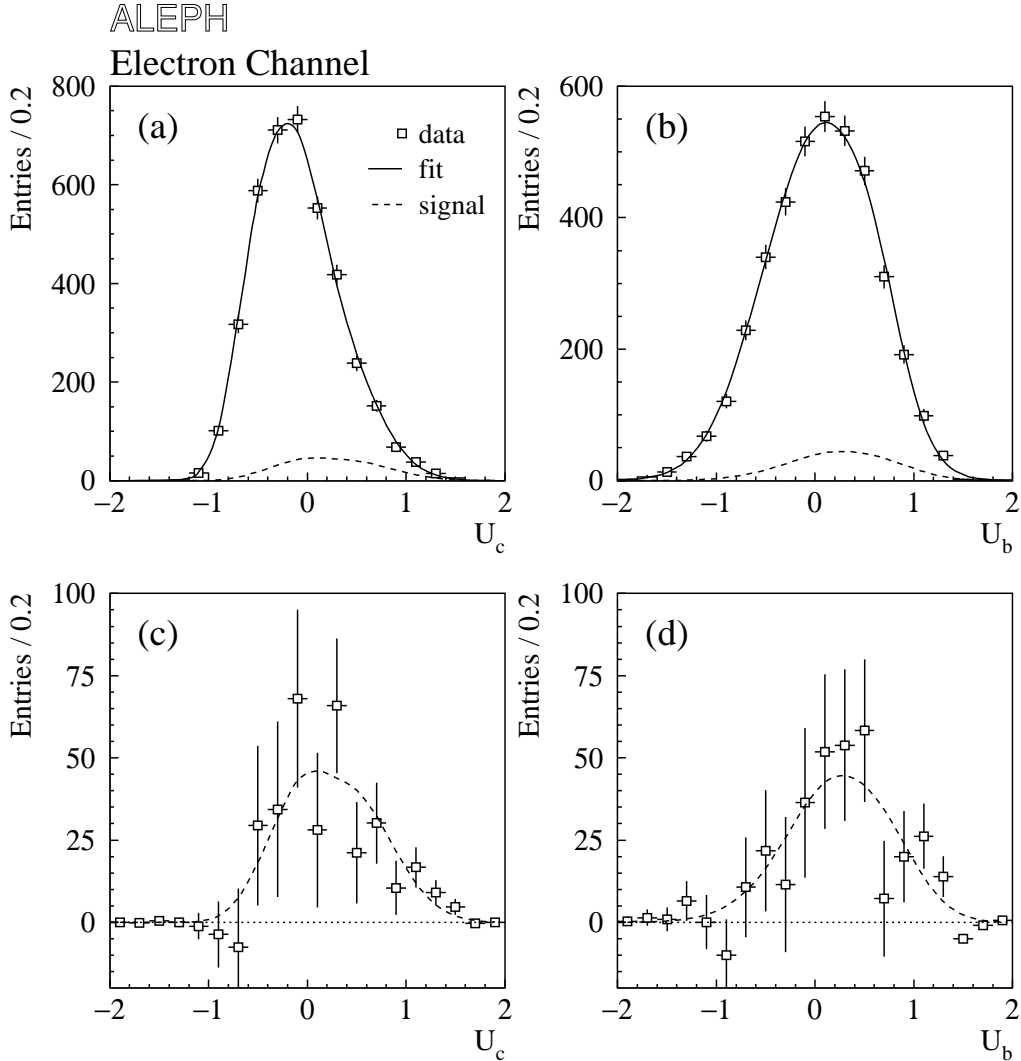}}%
\end{center}
\caption{\label{f:ftaue} 
Projections of the fit to the data in the electron channel
of the $\Dstaunu$ analysis:
(a) $\Uc$ distribution;
(b) $\Ub$ distribution.
The data are shown by the squares with error bars.
The solid curves are the fitted distributions, while
the dashed curves show the signal contributions.
The same distributions are shown in (c) and (d),
after subtraction of the fitted backgrounds.}
\end{figure}

\begin{figure}[p]
\begin{center}
\mbox{%
\includegraphics[height=150mm]{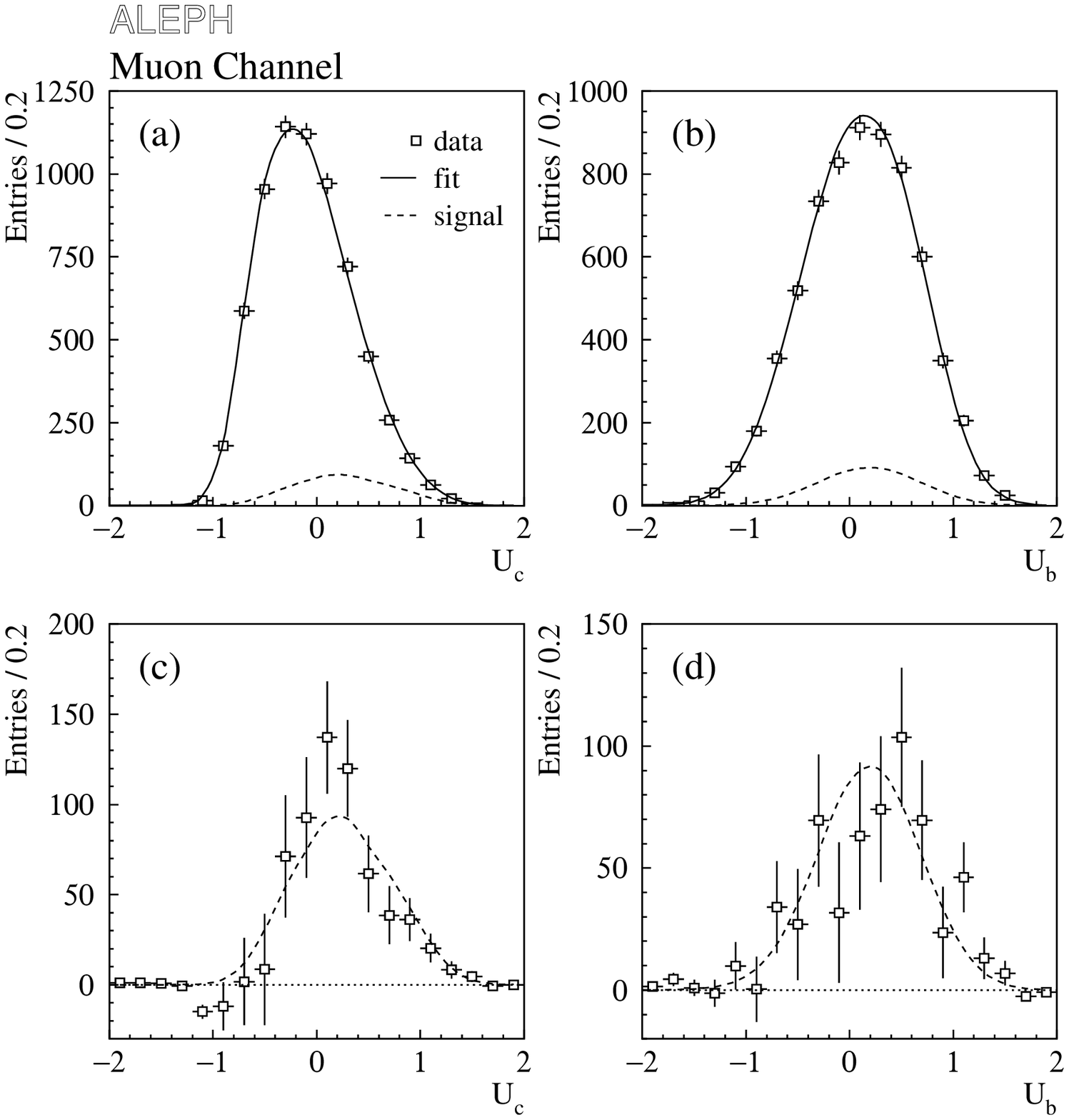}}%
\end{center}
\caption{\label{f:ftaum} 
Projections of the fit to the data in the muon channel
of the $\Dstaunu$ analysis:
(a) $\Uc$ distribution;
(b) $\Ub$ distribution.
The data are shown by the squares with error bars.
The solid curves are the fitted distributions, while
the dashed curves show the signal contributions.
The same distributions are shown in (c) and (d),
after subtraction of the fitted backgrounds.}
\end{figure}

The projections of the two-dimensional fits to the data are shown in 
Figs.~\ref{f:ftaue} and \ref{f:ftaum}.
The fitted numbers of signal and background events are listed
in Table~\ref{t:tauevts}.
The fitted branching fractions from the electron and muon channels are
$B(\Dstaunu) = [5.86 \pm 1.18{\rm (stat)}]\%$ and 
$[5.78 \pm 0.85{\rm (stat)}]\%$,
respectively. 

The goodness of fit is evaluated by comparing the likelihood value
obtained from each fit to the data
with the distribution of likelihoods obtained from fits to
many toy Monte Carlo samples.
The fitting functions are used to define
the parent distributions in generating these samples.
The fit confidence levels are calculated to be
83\% in the electron channel and
81\% in the muon channel;
the $\Ub$~versus $\Uc$ distributions in the data
are well described by the fitting functions.
\begin{table}[t]
\caption{{\label{t:tauevts}} 
Fitted numbers of signal and background events in data in the
$\Dstaunu$ analysis.
Only the statistical uncertainties from the fit are shown.}
\begin{center}
\vspace{2mm}
\begin{tabular}{l@{\hspace{10mm}}r@{$\,$}c@{$\,$}r@{\hspace{10mm}}r@{$\,$}c@{$\,$}r}\hline
Component           & \multicolumn{3}{l}{e channel} & \multicolumn{3}{c}{$\mu$ channel} \\ \hline
Signal              &  306 & $\pm$ &  62  &   575 & $\pm$ &  84 \\
uds background      &  111 & $\pm$ &  56  &   455 & $\pm$ & 139 \\
$\ccbar$ background & 2310 & $\pm$ & 101  &  3750 & $\pm$ & 182 \\
$\bbbar$ background & 1228 & $\pm$ &  56  &  1857 & $\pm$ &  74 \\ \hline
\end{tabular}                     
\end{center}
\end{table}
The same projections are also shown in
Figs.~\ref{f:ftaue} and \ref{f:ftaum}\ 
after subtraction of the fitted background components.
The distribution of the excess events is consistent in shape
with the Monte Carlo prediction for the signal.

\begin{figure}[p]
\begin{center}
\mbox{%
\includegraphics[height=150mm]{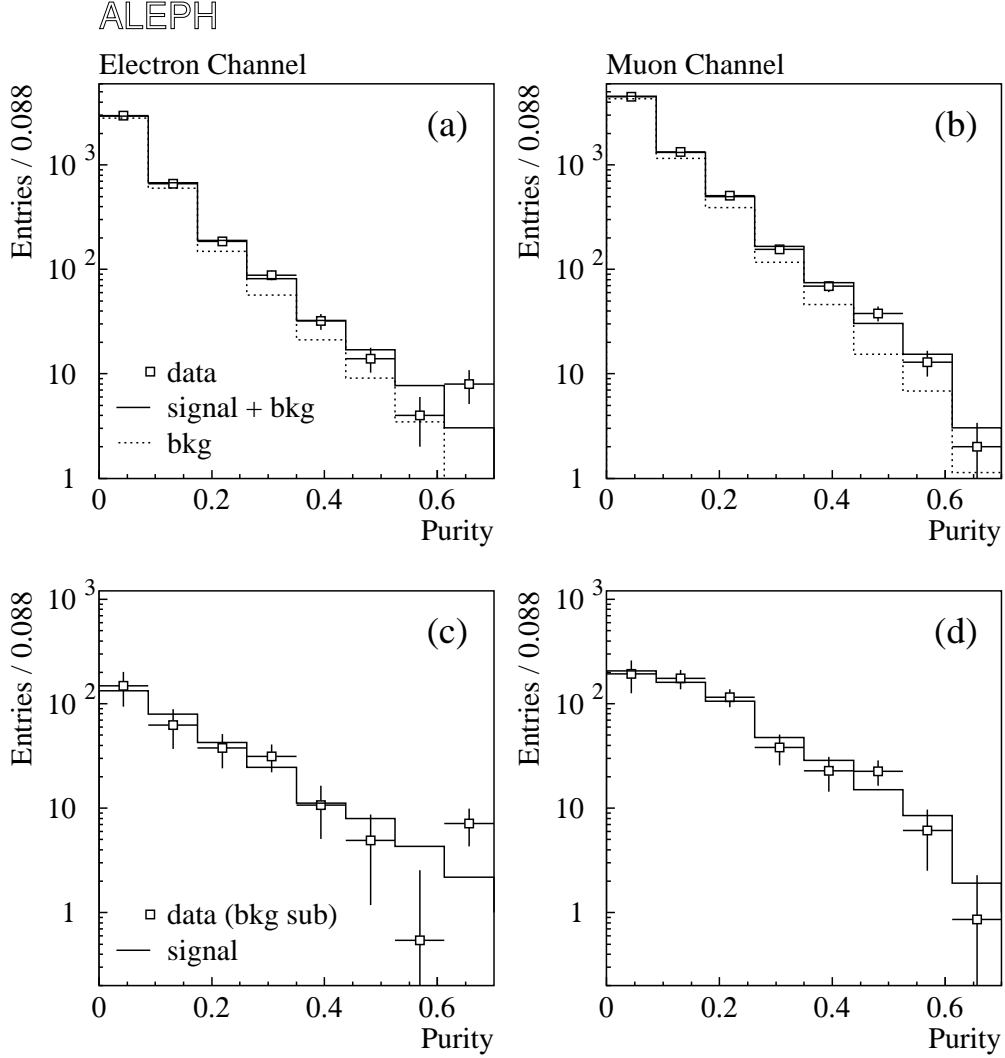}}%
\end{center}
\caption{\label{f:puretau} 
Distributions of purity $P$, in the 
(a) electron channel and
(b) muon channel of the $\Dstaunu$ analysis.
The data are shown by the squares with error bars.
The histograms show the Monte Carlo distributions
for the background (dotted) and for
the signal plus background (solid).
The Monte Carlo histograms are normalized according
to the results of the fits to the data.
The plots in (c) and (d) show the same
distributions after subtraction of the fitted backgrounds;
the squares are the data and the histograms
are the Monte Carlo signal distributions.
The error bars in all four plots
reflect the statistical errors in the data.}
\end{figure}

It is desirable to visualize the results of the
two-dimensional fits in one-dimensional distributions.
For this purpose, the purity $P$ of each event is calculated.
The purity of event $i$ is defined as
\[ P_i = \frac{S(\Uargs)}{S(\Uargs)+B(\Uargs)}\ , \]
where $\Uci$ and $\Ubi$ are the values of the
discriminant variables for event $i$,
and $S$ and $B$ represent the distributions
of the signal and background,
normalized according to the results of the fit.
The distributions of $P$ are quite different
for signal and background events.
The purity distributions from the electron and
muon channels are shown in Figs.~\ref{f:puretau}a and b;
the background-subtracted distributions are given in
Figs.~\ref{f:puretau}c and d.
These figures show that the shapes of the observed
purity distributions agree with the expectations.
In particular,
the shape of the data distribution in each channel 
after background subtraction
is consistent with the simulated signal
and inconsistent with the simulated background.

% ----------------------------------------------------------------------
\section{\boldmath$\Dsmunu$ analysis}
\label{s:munu}%
The second analysis~\cite{john} is optimized to select
$\ee\To\ccbar$ events containing the decay $\Dsmunu$.

\subsection{Event selection}
The same preselection of hadronic Z decays is performed
as in the $\taunu$ analysis.
A cut on the number of reconstructed charged tracks
coming from the area of the interaction point is applied
to reduce the background from dilepton events.
The event thrust axis is required to satisfy
$|{\cos\theta_{\rm thrust}}| < 0.8$.
A loose muon identification algorithm~\cite{john},
requiring either muon chamber hits or a muon-like digital pattern
in the HCAL, is used to select muon candidates.
If more than one muon is present in the event,
the one with highest momentum is selected.

A kinematic fit is then performed in order to improve the
resolution on the missing momentum, assumed to arise from
the undetected neutrino from $\Dsmunu$.
In this fit, the missing mass is constrained to be zero
and the $\Ds$ displacement direction 
(from the event primary vertex to the $\Ds$ decay vertex,
i.e., an unknown point on the muon track)
is constrained to be parallel to
the $\Ds$ momentum direction.
This constraint is equivalent to requiring that the reconstructed
neutrino direction be parallel to a plane containing the 
primary vertex and the muon track.
The energies of the reconstructed
charged and neutral particles are varied in the fit,
while their directions are held constant.
The muon track impact parameters and the coordinates of the event 
primary vertex are also allowed to vary within their uncertainties.
The kinematic fit improves the resolution on the neutrino energy 
from $5.7\gev$ to $2.8\gev$
and on the neutrino direction from $8.9^{\circ}$ to $6.2^{\circ}$.

The $\bbbar$ background and some of the $\ccbar$ background
are further reduced by cuts on
the track pseudorapidities and impact parameters,
as described in Section~\ref{s:taunu:sel}.
Finally, a hard cut is made on the fitted energy of the $\Ds$ candidate,
$E_{\mu} + E_{\nu} > 28\gev$.
This procedure selects 7164 events in the data.

\subsection{Linear discriminant analysis}
\begin{figure}[p]
\begin{center}
\mbox{\includegraphics[height=150mm]{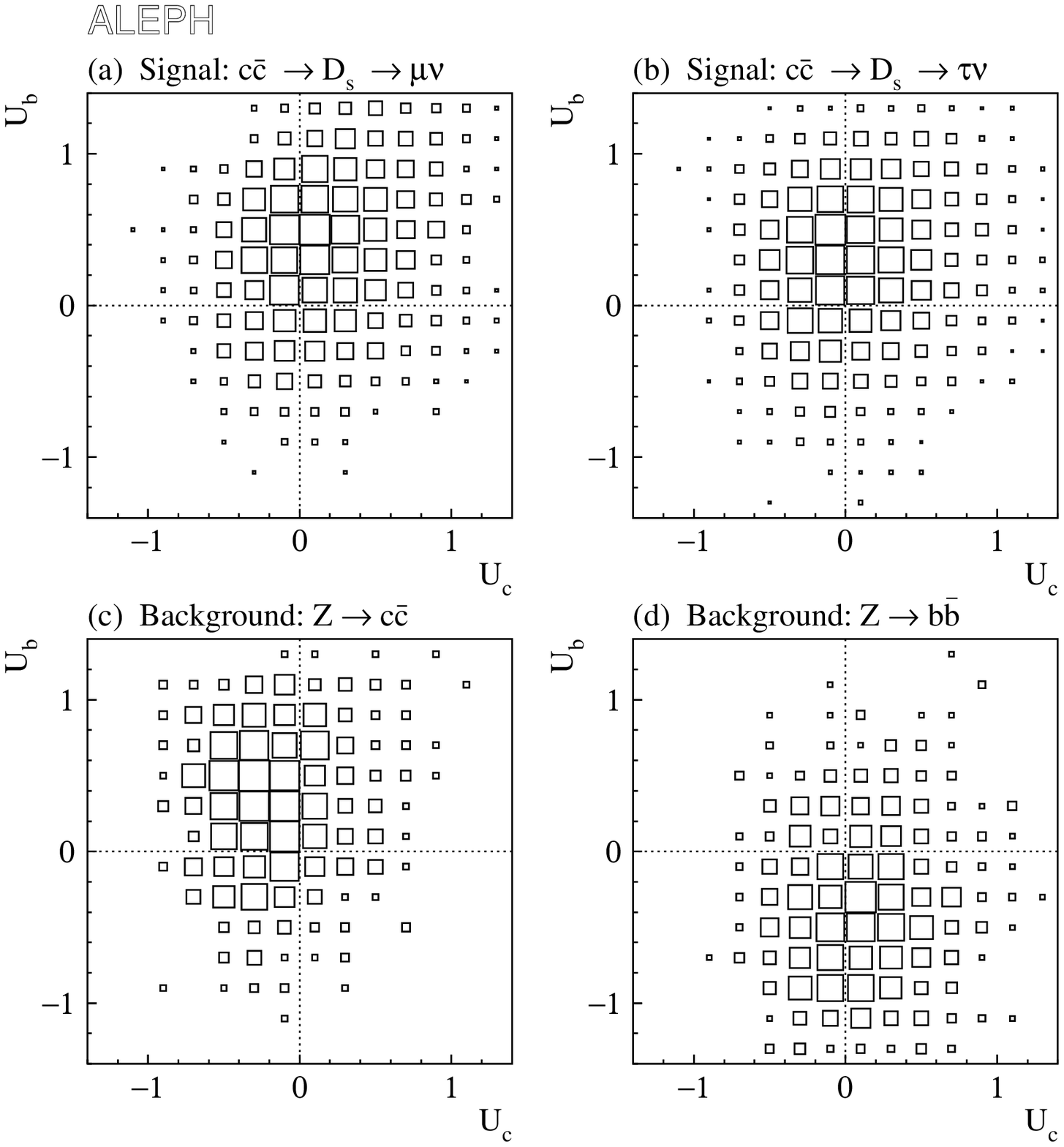}}
\end{center}
\caption{\label{f:umunu}
Monte Carlo $\Ub$~versus $\Uc$ distributions in the $\Dsmunu$ analysis, 
after all cuts:
(a) signal, $\ccbar\To\Dsmunu$;
(b) $\ccbar\To\Dstaunu$, which is also included in the signal component
of the fitting function;
(c) $\ccbar$ background;
(d) $\bbbar$ background.
The distributions are shown with arbitrary normalizations.}
\end{figure}

Linear discriminant variables $\Uc$ and $\Ub$ are
optimized to distinguish $\ccbar\To\Dsmunu$ signal events from
$\ccbar$ and $\bbbar$ backgrounds, respectively.
Each of the linear discriminant variables is a combination
of seven event variables.
The variables with the greatest discrimination
power include the muon momentum, the fitted neutrino momentum,
and several b- and c-tag neural network outputs.
The $\Ub$~versus $\Uc$ distributions for signal and background
events are shown in Fig.~\ref{f:umunu}.

\subsection{Results}
The numbers of signal and background events are
extracted from the data by means of 
a maximum likelihood fit to the 
$(\Uc,\Ub,\Mmunu)$ distribution,
where $\Mmunu$ is the invariant mass of the $\Ds$ candidate
after the kinematic fit.
The three-dimensional region
$-0.6 < \Uc < 1.0$, $-0.8 < \Ub < 1.4$, and $\Mmunu < 5\gevcc$
is divided into $6 \times 6 \times 11$ bins in the fit.
The fitting functions are constructed from simulated events
and a procedure is applied to smooth the distributions
to reduce the bias that results from limited Monte Carlo
statistics near the edges of the space.

As in the $\Dstaunu$ analysis,
the signal component consists of eight contributions
from $\Ds$ and $\Dp$ decays to $\taunu$ and $\munu$
in $\ccbar$ and $\bbbar$ events;
the relative normalizations of these contributions are again fixed
according to the Standard Model expectations (Table~\ref{t:musig}).
Although the $\munu$ invariant mass provides some separation
between $\Dsmunu$ decays and the backgrounds,
the $\Dsmunu$ decay mode comprises only 30\% of the signal
and cannot be clearly isolated in a fit to the
$(\Uc,\Ub,\Mmunu)$ distribution.
With the Standard Model constraint
$B(\Dsmunu)/B(\Dstaunu) = 0.103$ imposed in the fit,
this analysis nevertheless yields additional information on $\fDs$.
(Fits in which this constraint is not imposed are described
in Section~\ref{ss:checks}.)
\begin{table}[t]
\begin{center}
\caption{Relative normalizations of the contributions 
to the signal in the $\Dsmunu$ analysis.}{\label{t:musig}}
\vspace{2mm}
\begin{tabular}{lr}                                       \hline
Source         & Fraction (\%)   \\ \hline
$\ccbar\To\Dstaunu$   &  49.6\hS \\
$\ccbar\To\Dsmunu$    &  26.7\hS \\
$\bbbar\To\Dstaunu$   &  12.6\hS \\
$\bbbar\To\Dsmunu$    &   3.5\hS \\
$\ccbar\To\Dptaunu$   &   2.2\hS \\
$\ccbar\To\Dpmunu$    &   4.3\hS \\
$\bbbar\To\Dptaunu$   &   0.6\hS \\
$\bbbar\To\Dpmunu$    &   0.5\hS \\ \hline
Total                 & 100.0\hS \\ \hline
\end{tabular}
\end{center}
\end{table}

\begin{figure}[p]
\begin{center}
\mbox{%
\includegraphics[height=150mm, bb=0 0 265 567]{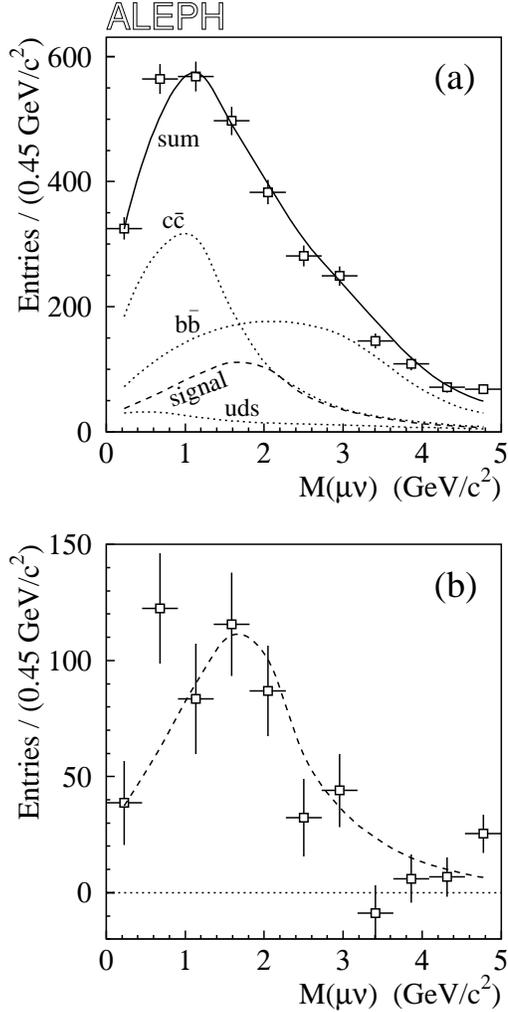}}
\end{center}
\caption{\label{f:datafit} 
(a) $\Mmunu$ projection of the fit to the data
in the $\Dsmunu$ analysis.  
The data are shown by the squares with error bars.
The solid curve is the fitted distribution.
The dashed curve shows the signal contribution,
while the dotted curves indicate the three background contributions.
The data and fitted signal distributions are shown in (b),
after subtraction of the fitted backgrounds.}
\end{figure}

The fitted numbers of events are
$553 \pm 93$ signal events,
$166 \pm 47$ $\uds$ background events, 
$1251 \pm 71$ $\ccbar$ background events, and 
$1291 \pm 62$ $\bbbar$ background events.
Figure~\ref{f:datafit} shows the fit
projection on the $\Mmunu$ axis.
After correction for a bias of +8.2\% 
in the smoothing and fitting procedure,
the fit result corresponds to
$B(\Dsmunu)= [0.68 \pm 0.11({\rm stat})]\%$.
The goodness of fit is characterized by a 
confidence level of 69\%.

\begin{figure}[p]
\begin{center}
\mbox{%
\includegraphics[height=150mm, bb=0 0 265 567]{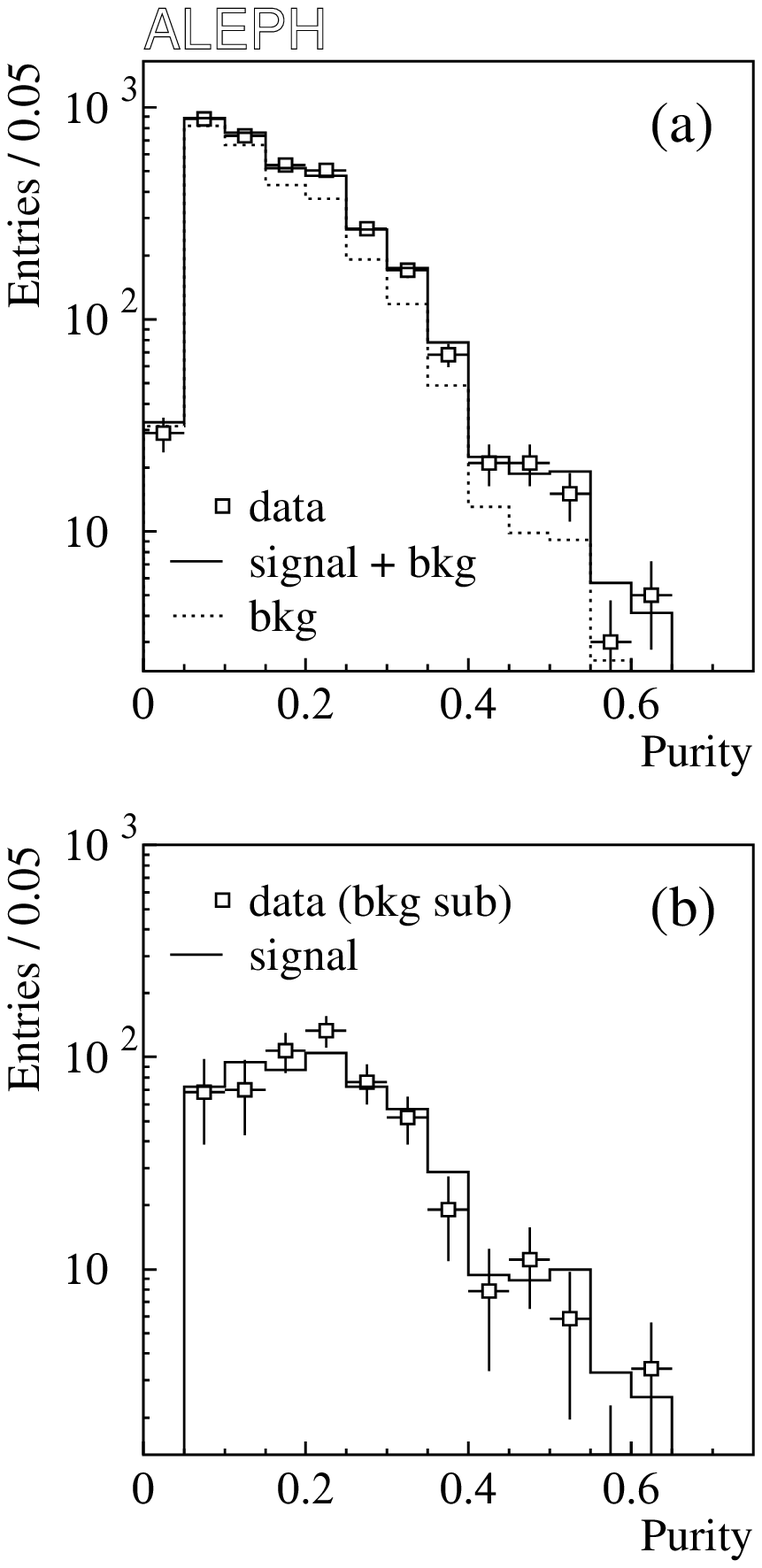}}
\end{center}
\caption{\label{f:puremu} 
(a) Distribution of purity $P$ in the $\Dsmunu$ analysis.
The data are shown by the squares with error bars.
The histograms show the Monte Carlo distributions
for the background (dotted) and for
the signal plus background (solid).
The Monte Carlo histograms are normalized according
to the results of the fit to the data.
(b) Purity distribution after subtraction of the 
fitted backgrounds;
the squares are the data and the histogram
is the Monte Carlo signal distribution.
The errors in both plots
reflect the statistical errors in the data.}
\end{figure}
The purity distribution from the $\Dsmunu$ analysis is
shown in Fig.~\ref{f:puremu}.
The purity is defined as in Section~\ref{s:taunu:results},
except in this case the signal and background distributions $S$ and $B$
are binned functions of $\Uc$, $\Ub$, and $\Mmunu$.
Again the signal and background events have distinct
distributions,
and the shape of the background-subtracted data distribution agrees with
that of the simulated signal.

% ----------------------------------------------------------------------
\section{Systematic errors}
\label{s:syst}%
\subsection{Evaluation of systematic uncertainties}
In both analyses, 
the shapes of the signal and background fitting functions 
are taken from simulated events.
The physics parameters that can affect these shapes are
taken into account 
in the systematic uncertainties on the branching fractions.

The sources of systematic error are summarized in Table~\ref{t:syst}.
The magnitudes of the systematic errors are estimated by
measuring the branching fractions from a Monte Carlo sample,
reweighting the Monte Carlo events to simulate a variation in a parameter,
then measuring the branching fractions again.
In each case the quoted systematic uncertainty is taken to be the quadratic sum
of the observed shift and the statistical uncertainty on the shift.
\begin{table}[t]
\caption{\label{t:syst}
Relative systematic uncertainties on $B(\Dslnu)$.}
\begin{center}
\begin{tabular}{l@{\hspace{6mm}}r@{\hspace{8mm}}r@{\hspace{8mm}}r}\hline
                         & \multicolumn{3}{c}{Uncertainty $(\%)$}   \\
Source                   & $\Dstaunu$ (e)& $\Dstaunu$ ($\mu$) & $\Dsmunu$ \\ \hline
Charm hadron production            & 27.0\hS & 22.8\hS & 19.6\hs \\ 
c fragmentation                    & 16.7\hS & 14.8\hS & 12.1\hs \\ 
b fragmentation                    &  9.0\hS &  5.9\hS &  4.1\hs \\ 
Lepton spectra in b and c decays   &  5.5\hS &  3.6\hS &  3.1\hs \\ 
Lepton rates in $\bbbar$ events    &  0.3\hS &  0.1\hS &  1.7\hs \\  
Detector resolution                & 13.6\hS & 12.7\hS &  4.0\hs \\ 
Monte Carlo statistics             &  7.2\hS &  6.0\hS &  8.8\hs \\ 
Bias in fitting procedure          &  ---\hS &  ---\hS &  4.1\hs \\
$\Dp\To\Kzbar\mu^+\nu$ form factor &  0.5\hS &  0.5\hS &  1.5\hs \\ 
$\fDs/\fDp$                        &  0.2\hS &  0.7\hS &  1.4\hs \\ \hline
Total                              & 36.8\hS & 31.4\hS & 26.0\hs \\ \hline
\end{tabular}
\end{center}
\end{table}

The uncertainties on the production rates of $\Dp$, ${\rm D}^0$,
$\Ds$, and charm baryons from~\cite{lepewwg,ccount} are taken
into account, including correlations.
The $\Ds$ production rates have large uncertainties
due mainly to the large uncertainty on the absolute scale of
branching fractions for hadronic $\Ds$ decays;
these parameters give the largest contributions to the systematic error
on the leptonic branching fractions because they directly govern
the number of produced $\Ds$ mesons in the data sample.

The Monte Carlo events are weighted to match the mean scaled energies 
$\langle x_E({\rm c}) \rangle = 0.484 \pm 0.008$ and
$\langle x_E({\rm b}) \rangle = 0.702 \pm 0.008$ 
of charm and b hadrons given in~\cite{LEPHF98};
the uncertainties on those fragmentation parameters 
are propagated to the leptonic branching fractions.

The shapes of the lepton energy spectra in b and c decays
are varied according to the prescription given in~\cite{LEPHF94}.
The ${\rm b}\To\Ell$ and ${\rm b}\To{\rm c}\To\bar{\Ell}$
fractions and uncertainties are taken from~\cite{lepewwg}.

The detector resolution on missing energy and momentum is
dominated by the errors on neutral particle energies.
The energy resolutions for neutral energy flow particles
are studied in samples of hadronic $\Z$ decays in data and Monte Carlo.
The electromagnetic and hadronic energy resolution parameters 
for each sample are estimated by means of a global fit in which
the total energy in each event is constrained to $\sqrt{s}$
and the total momentum is constrained to zero.
Although the four-momentum carried away by neutrinos and
objects at low angles is not taken into account in this procedure,
a useful comparison of the energy resolutions 
in data and Monte Carlo can nevertheless be made.
A discrepancy of less than 4\% in the neutral particle
energy resolution is observed,
and the effect on the branching fractions is estimated by 
further smearing 
the neutral energies in the Monte Carlo events 
that are used to build the fitting functions.

The Monte Carlo statistical uncertainty is estimated by 
generating many toy Monte Carlo samples,
using the fitting functions obtained from the full simulation
as the parent distributions.
This procedure reveals a $+8.2\%$ bias in the smoothing and
fitting procedure in the $\Dsmunu$ analysis,
and a systematic uncertainty equal to half the bias is assigned.
As a cross check,
a fit to the data in this channel
is also performed with an algorithm in which 
the statistical fluctuations in the Monte Carlo distributions
are taken into account~\cite{hmcmll};
the fitted number of signal events is within $2\%$
of that obtained with the standard program, after bias corrections.

Other small effects considered are the form factor in
$\Dp\To\Kzbar\mu^+\nu$ decays~\cite{CLEO_FF}
(an important source of background)
and the ratio of decay constants $\fDs/\fDp$
from~\cite{draper}.

\subsection{\boldmath Cross check with $\Dsphipi$ decays}
A study of $\Dsphipi$ decays is performed as a cross check
of the signal efficiencies in the $\Dslnu$ analyses.
Candidate $\Dsphipi$ decays with $\phi\To\Kp\Km$
are first selected from the full $\ee\To\qqbar$ sample.
The $dE/dx$ measurements in the TPC are used to 
discriminate between pions and kaons.
Cuts are also applied on the kaon and pion momenta,
the $\Kp\Km$ invariant mass, and
the decay angle of the $\phi$.
The same selection is applied to simulated $\qqbar$ events 
containing $\Dsphipi$ decays.

The $\KKpi$ candidates are divided into seven bins
of $x_E = E_{\KKpi}/E_{\rm beam}$,
from 0.3 to 1.
The number of $\Ds$ mesons in each $x_E$ bin in the data
is evaluated by means of a fit to
the reconstructed $\KKpi$ invariant mass distribution
for the candidates in that bin.
In this fit the signal is described as the sum of two
Gaussian functions,
and a second pair of Gaussians is included for the
$\Dp\To\phi\pi$ contribution;
a polynomial function is used for the background.

In order to measure the efficiencies of the $\Dslnu$ selections,
the pion in each $\KKpi$ combination
is treated as the lepton candidate; 
the kaons are removed from the event and correspond
to the missing neutrino(s).
After the selection cuts are made,
the number of $\Ds$ mesons in each $x_E$ bin
is again extracted from the $\KKpi$ invariant mass distribution.
The selection efficiency in each bin is then evaluated
in data and Monte Carlo.
Some cuts, most notably the lepton identification cuts,
cannot be studied with this method.
The $\Dstaunu$ and $\Dsmunu$ analyses are checked separately.

\begin{figure}[p]
\begin{center}
\mbox{%
\includegraphics[height=82.5mm, bb=0 255 312 567]{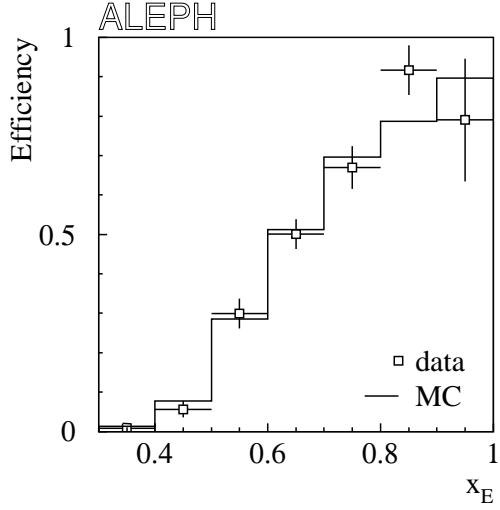}}
\end{center}
\caption{\label{f:efftau}%
Efficiency versus $x_E$ in data (squares with error bars)
and Monte Carlo (histogram)
for the $\Dstaunu$ selection cuts listed in the text,
as measured from $\Dsphipi$ decays.
The cut on the fitted $\Ds$ energy,
$E_{\Ds} > 22.5\gev$,
corresponds to $x_E \gequ 0.49$.}
\end{figure}
\begin{figure}[p]
\begin{center}
\mbox{%
\includegraphics[height=82.5mm, bb=0 255 312 567]{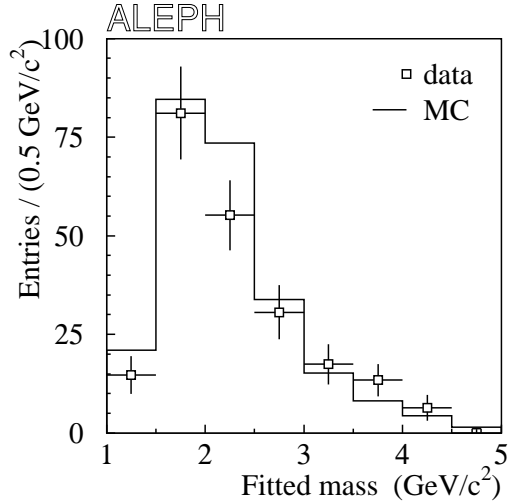}}
\end{center}
\caption{\label{f:phimass}%
Background-corrected distribution of
the $\Ds$ mass 
(calculated from the $\pi$ and the missing $\Kp\Km$)
after the kinematic fit,
in the $\phi\pi$ cross check of the $\Dsmunu$ analysis.
The data are shown by the squares with error bars,
while the histogram shows the Monte Carlo distribution.
The number of $\Ds$ mesons in each mass bin in the data
is obtained from a fit to the corresponding distribution of
the fully reconstructed $\KKpi$ invariant mass.}
\end{figure}
The following aspects of the $\Dstaunu$ analysis
are included in the $\phi\pi$ cross check:
the cut on missing energy,
the $\bbbar$ and dilepton rejection cuts,
the kinematic fit,
and the cut on the fitted $\Ds$ energy.
With the kaon candidates excluded from the event,
these operations can be accurately duplicated in the $\phi\pi$ events.
The overall efficiency of these cuts as a function of $x_E$,
in data and Monte Carlo,
is plotted in Fig.~\ref{f:efftau}.
The signal efficiency (apart from lepton identification)
is verified to be correctly simulated within the
statistical uncertainty of the test (7\% relative uncertainty).

A similar $\phi\pi$ cross check is performed for the
$\Dsmunu$ analysis.
The main differences with respect to the $\Dstaunu$ case
are (1) the fitted primary vertex
and the impact parameters of the pion (serving as the muon) 
are now involved in the kinematic fit, and
(2) it is possible to check the resolution on the
reconstructed $\Ds$ mass.
For purposes of this test, the missing mass
is constrained in the kinematic fit to equal
the reconstructed $\Kp\Km$ invariant mass
on an event-by-event basis.
The measured efficiencies in data and Monte Carlo
are again found to be in agreement.
The distributions of reconstructed $\Ds$ mass 
in data and Monte Carlo are plotted in Fig.~\ref{f:phimass};
the simulated resolution is found to be consistent
with that observed in the data.

\subsection{Additional checks}
\label{ss:checks}%
Additional checks, described in this section, are performed
in order to further investigate the accuracy of the Monte Carlo
simulation.

The fitted numbers of 
$\uds$, $\ccbar$, and $\bbbar$ background events
are compared with the Monte Carlo predictions
in Table~\ref{t:bkg}.
The normalization of the predicted numbers is based
on the total number of produced $\ee\To\qqbar$ events
in the data sample.
The predictions for $\Dsmunu$ are corrected for biases
in the fitting procedure.
The data and Monte Carlo are in agreement.
A $\chi^2$ value is computed to characterize this agreement
in the $\Dsmunu$ analysis, taking into account the correlations
in the statistical and systematic errors on the normalizations
of the three backgrounds.
The result is $\chi^2 = 0.74$ for three degrees of freedom
(${\rm CL} = 87\%$).
\begin{table}[t]
\caption{{\label{t:bkg}}Fitted background events relative to
the expected numbers 
in the two channels of the $\Dstaunu$ analysis and in the $\Dsmunu$ analysis.
The first uncertainty given in each case is statistical
and the second systematic.}
\begin{center}
\begin{tabular}{l@{\hspace{7mm}}ccc}\hline
          &\multicolumn{3}{c}{Fitted / Expected Background} \\
Background & $\Dstaunu$ (e)       & $\Dstaunu$ ($\mu$)   & $\Dsmunu$            \\ \hline
$\uds$   & $0.81\pm0.40\pm0.58$ & $1.00\pm0.31\pm0.42$ & $0.84\pm0.22\pm0.17$ \\ 
$\ccbar$ & $0.92\pm0.04\pm0.06$ & $0.93\pm0.04\pm0.07$ & $0.96\pm0.05\pm0.08$ \\ 
\rule{0pt}{4.2mm}%
$\bbbar$ & $0.92\pm0.05\pm0.05$ & $1.06\pm0.04\pm0.05$ & $0.98\pm0.05\pm0.06$ \\ \hline
\end{tabular}
\end{center}
\end{table}

The distributions of the event variables
used to construct the three sets of linear discriminant
variables $\Uc$ and $\Ub$ are compared in data and Monte Carlo to ensure
that the simulation is reliable.
The Monte Carlo plots are normalized according to the results of the fits to the data.
No significant discrepancies are found.

Another cross check on the accuracy of the simulation is made by
comparing the distributions of certain quantities 
in data and Monte Carlo 
in different regions of the $(\Uc,\Ub)$ space.
In each analysis, four disjoint regions in that space are defined 
with different relative abundances of the signal and the three backgrounds.
The Monte Carlo normalization is again based on the results of the fits to the data.
The quantities studied 
in the $\Dstaunu$ analysis include
the fitted $\Ds$ momentum,
the angle between the lepton and the $\Ds$ momentum in the $\Ds$ rest frame,
and b tag variables;
in the $\Dsmunu$ analysis
the missing energy before the kinematic fit,
the $\munu$ invariant mass after the kinematic fit,
and the fitted proper $\Ds$ decay length
were considered.
The simulation is found to be consistent with the data
in all cases and in all regions of the $(\Uc,\Ub)$ space.
This check demonstrates that the properties of the
backgrounds are correctly reproduced in the Monte Carlo.

The fit to the data in the $\Dsmunu$ analysis was
repeated without the Standard Model constraint
on $B(\Dsmunu)/B(\Dstaunu)$.
For this calculation,
the signal contributions from $\Ds$ and $\Dp$ decays to $\munu$
were allowed to vary separately from the $\taunu$ contributions.
The fitted contribution from $\munu$ relative to $\taunu$
is found to be somewhat smaller than expected;
Monte Carlo experiments show that the unconstrained fit
is expected to give a $\munu/\taunu$ ratio smaller than
the observed value 5\% of the time
when the events are generated according to the Standard Model ratio.
The fitted $\munu/\taunu$ ratio is not sensitive to the
assumed charm production rates or fragmentation parameters.

Increased sensitivity to $B(\Dsmunu)$ is achieved if
the $\Dsmunu$ analysis is coupled to
the statistically independent electron channel 
of the $\Dstaunu$ analysis.
The fit described in the preceding paragraph is repeated
with a modified likelihood function in which
$B(\Dstaunu)$ is constrained according to the value
and uncertainty obtained from the electron channel;
after correction for the calculated $-1\%$ and $+14\%$ biases,
respectively,
the results are
\begin{eqnarray*}
B(\Dsmunu)  &=& (0.47 \pm 0.25) \% \\
B(\Dstaunu) &=& (6.8 \pm 1.0) \% \ ,
\end{eqnarray*}
where the uncertainties are statistical only.

% ----------------------------------------------------------------------
\section{Conclusions}
\label{s:conc}%
Leptonic decays of the $\Ds$ meson have been studied in
a sample of four million hadronic $\Z$ decays.
The $\Dstaunu$ analysis gives
\begin{eqnarray*}
B(\Dstaunu) &=& \rlap{\mbox{\hspace{48mm}(electron channel)}}
                (5.86 \pm 1.18 \pm 2.16)\% \ , \\
            &=& \rlap{\mbox{\hspace{48mm}(muon channel)}}
                (5.78 \pm 0.85 \pm 1.81)\% \ ,
\end{eqnarray*}
where the measured signal includes some $\Dsmunu$ and
$\Dplnu$ decays.
In the Standard Model,
these branching fractions correspond to the following values of the
decay constant $\fDs$:
\begin{eqnarray*}
\fDs        &=& \rlap{\mbox{\hspace{56mm}(electron channel)}}
                (275 \pm 28 \pm 51)\mev \ , \\
            &=& \rlap{\mbox{\hspace{56mm}(muon channel)}}
                (273 \pm 20 \pm 43)\mev \ .
\end{eqnarray*}

These results are combined, 
taking into account the common systematic errors~\cite{combine},
to obtain
\begin{eqnarray*}
B(\Dstaunu) &=& (5.79 \pm 0.77 \pm 1.84) \% \\
\fDs        &=& (273 \pm 18 \pm 43)\mev \ .
\end{eqnarray*}
The combination has $\chi^2 = 0.0024$ for 1 degree of freedom (${\rm CL} = 96\%$).
This measurement may be expressed as
\[ f({\rm c}\To\Ds) B(\Dstaunu) = (6.77 \pm 0.90 \pm 1.49) \times 10^{-3} \ , \]
where $f({\rm c}\To\Ds)$ is the $\Ds$ production rate per hemisphere in
$\Z\To\ccbar$ events;
the relative uncertainty on this product is smaller than that on $B(\Dstaunu)$
due to the reduced dependence on the assumed $\Ds$ production rate.

The analysis optimized for $\Dsmunu$ is also sensitive to $\Dstaunu$ decays;
a measurement of the combined signal yields
\[ B(\Dsmunu) = (0.68 \pm 0.11 \pm 0.18)\% \]
when the ratios of the signal components are fixed to their Standard Model values.
This measurement corresponds to
\[ \fDs = (291 \pm 25 \pm 38)\mev \ . \]

Finally, the three $\fDs$ measurements 
(two from $\taunu$ and one from $\munu$)
are combined.
The correlation coefficient $k = +0.43 \pm 0.34$ of the
statistical errors in the $\Dsmunu$ analysis and the
muon channel of the $\Dstaunu$ analysis
is evaluated by dividing the data into ten subsamples.
The correlations between the systematic errors are also
taken into account.
The combined decay constant is
\[ \fDs = (285 \pm 19 \pm 40) \mev \]
with $\chi^2 = 0.51$ for 2 degrees of freedom (${\rm CL} = 78\%$).

This result is consistent with
recent measurements~\cite{rpp,beatrice,opal} 
and with the lattice QCD prediction,
$\fDs = 255 \pm 30 \mev$~\cite{bernard}.

% ----------------------------------------------------------------------
\section*{Acknowledgements}
We wish to thank our colleagues in the CERN accelerator divisions
for the successful operation of LEP.
We are indebted to the engineers and technicians in all our
institutions for their contribution to the excellent performance
of ALEPH.
Those of us from nonmember countries thank CERN for its hospitality.

% ----------------------------------------------------------------------

\end{document}